\documentclass{article}
\usepackage{amssymb}
\usepackage{amsmath}

\begin{document}

\begin{center}
\textbf{Nature of Electric and Magnetic Fields; How the Fields
Transform\bigskip \medskip }

Tomislav Ivezi\'{c}$\bigskip $

\textit{Ru%
\mbox
 {\it{d}\hspace{-.15em}\rule[1.25ex]{.2em}{.04ex}\hspace{-.05em}}er Bo\v{s}%
kovi\'{c} Institute, P.O.B. 180, 10002 Zagreb, Croatia}

\textit{ivezic@irb.hr}\bigskip \bigskip
\end{center}

\noindent In this paper the proofs are given that the electric and magnetic
fields are properly defined vectors on the four-dimensional (4D) spacetime
(the 4-vectors in the usual notation) and not the usual 3D fields. They are
the 4D geometric quantities (GQs). Furthermore, the proofs are presented
that under the mathematically correct Lorentz transformations (LT), e.g.,
the electric field vector transforms as any other vector transforms, i.e.,
again to the electric field vector; there is no mixing with the magnetic
field vector $B$, as in the usual transformations of the 3D fields.
Different derivations of these usual transformations of the 3D fields,
including those from some well-known textbooks, are discussed and objected.
This formulation with the 4D GQs is in a true agreement, independent of the
chosen inertial reference frame and of the chosen system of coordinates in
it, with experiments in electromagnetism, e.g., the motional emf. It is not
the case with the usual 3D formulation which agrees with experiments only if
the standard basis is used and for $\gamma \simeq 1$.\bigskip

\textit{In our living arena, the four-dimensional (4D) spacetime, physical
laws, e.g.},\textit{\ the Lorentz force law, are geometric, coordinate-free
relationships between the 4D geometric, coordinate-free quantities.}\bigskip
\bigskip

\bigskip \bigskip

\noindent \textbf{1. Introduction}\bigskip

\noindent It is generally accepted that the electric and magnetic fields are
the 3D vectors and that their transformations, e.g., equations (11.148) and
(11.149) in [1], are the mathematically correct Lorentz transformations (LT)
of these fields. In this paper the transformations of the 3D fields $\mathbf{%
E}$ and $\mathbf{B}$ will be called the \textquotedblleft
apparent\textquotedblright\ transformations (AT). The name is explained
below. According to the mentioned AT, the transformed 3D vector $\mathbf{E}%
^{\prime }$ is expressed by the mixture of the 3D vectors $\mathbf{E}$ and $%
\mathbf{B}$, equation (11.149) in [1]. In the usual covariant approaches,
e.g., [1], the AT for the components of $\mathbf{E}$ and $\mathbf{B}$ are
derived assuming that for the observers in an inertial frame, the $S$\
frame, these components are identified with the six independent components $%
F^{\alpha \beta }$ of the electromagnetic field tensor. These
identifications are%
\begin{equation}
E_{i}=F^{i0},\qquad B_{i}=(1/2)\varepsilon _{ijk}F_{kj}  \label{ieb}
\end{equation}%
(the indices $i$, $j$, $k$, $...=1,2,3$), equation (11.137) in [1], e.g., $%
E_{x}=E_{1}=F^{10}$. The components of the 3D fields $\mathbf{E}$ and $%
\mathbf{B}$ are written with lowered (generic) subscripts, since they are
not the spatial components of the 4D quantities. This refers to the
third-rank antisymmetric $\varepsilon $ tensor too. The super- and
subscripts are used only on the components of the 4D quantities. The 3D $%
\mathbf{E}$ and $\mathbf{B}$ are \emph{geometric quantities in the 3D space}
and they are constructed from these six independent components of $F^{\mu
\nu }$ and \emph{the unit 3D vectors }$\mathbf{i}$, $\mathbf{j}$, $\mathbf{k}
$, e.g., $\mathbf{E=}F^{10}\mathbf{i}+F^{20}\mathbf{j}+F^{30}\mathbf{k}$.
Observe that $F^{\alpha \beta }$ is not a tensor\ since $F^{\alpha \beta }$
are only components implicitly taken in the standard basis. The components
are coordinate quantities and they do not contain the whole information
about the physical quantity, since a basis of the spacetime is not included.
Then, it is supposed that the same identification of the components as in
equation (\ref{ieb}) holds for a relatively moving inertial frame $S^{\prime
}$, i.e., for the transformed components $E_{i}^{\prime }$ and $%
B_{i}^{\prime }$

\begin{equation}
E_{i}^{\prime }=F^{\prime i0},\quad B_{i}^{\prime }=(1/2c)\varepsilon
_{ijk}F_{kj}^{\prime }.  \label{eb2}
\end{equation}%
The same remark about the (generic) subscripts holds also here. The
components $F^{\alpha \beta }$ transform under the LT as, e.g.,%
\begin{equation}
F^{\prime 10}=F^{10},\ F^{\prime 20}=\gamma (F^{20}-\beta F^{21}),\
F^{\prime 30}=\gamma (F^{30}-\beta F^{31}),  \label{fe}
\end{equation}%
which yields (by equations (\ref{ieb}) and (\ref{eb2})) that
\begin{equation}
E_{1}^{\prime }=E_{1},\ E_{2}^{\prime }=\gamma (E_{2}-\beta cB_{3}),\
E_{3}^{\prime }=\gamma (E_{3}+\beta cB_{2}),  \label{ee}
\end{equation}%
what is equation (11.148) in [1]. Thus, in the usual covariant approaches,
e.g., [1], \emph{the AT of the components of} $\mathbf{E}$ \emph{and} $%
\mathbf{B}$ \emph{are derived assuming that they transform under the LT as
the components of} $F^{\alpha \beta }$ \emph{transform}.

However, there are several objections to the mathematical correctness of
such a procedure. Some of them are the following:

1) As seen, e.g., from section 3.1 in [2], such an identification of the
components of $\mathbf{E}$ and $\mathbf{B}$ with the components of $%
F^{\alpha \beta }$ is synchronization dependent and, particularly, it is
meaningless in the \textquotedblleft radio,\textquotedblright\
\textquotedblleft r\textquotedblright\ synchronization, i.e., in the $%
\left\{ r_{\mu }\right\} $ basis, see [3] and below.

2) The 3D vectors $\mathbf{E}$, $\mathbf{B}$ and $\mathbf{E}^{\prime }$, $%
\mathbf{B}^{\prime }$ are constructed in both frames in the same way, i.e.,
multiplying the components, e.g., $E_{x,y,z}$ and$\ E_{x,y,z}^{\prime }$ by
the unit 3D vectors $\mathbf{i}$, $\mathbf{j}$, $\mathbf{k}$ and $\mathbf{i}%
^{\prime }$, $\mathbf{j}^{\prime }$, $\mathbf{k}^{\prime }$, respectively.
This procedure gives the AT of the 3D vectors $\mathbf{E}$ and $\mathbf{B}$,
equation (11.149) in [1]. But, as seen from (\ref{fc}), \emph{the components}
$F^{\alpha \beta }$ \emph{are multiplied by the bivector basis} $\gamma
_{\alpha }\wedge \gamma _{\beta }$ \emph{and not by the unit 3D vectors. In
the 4D spacetime the unit 3D vectors are ill-defined algebraic quantities
and there are no LT, or some other transformations, that transform the unit
3D vectors} $\mathbf{i}$, $\mathbf{j}$, $\mathbf{k}$ \emph{into the unit 3D
vectors} $\mathbf{i}^{\prime }$, $\mathbf{j}^{\prime }$, $\mathbf{k}^{\prime
}$.

In [4], in section 12.3.2 under the title \textquotedblleft How the Fields
Transform,\textquotedblright\ the AT, equations (12.109), are derived using
the Lorentz contraction and the 3D fields. But, as shown, e.g., in [3] and
[5], the Lorentz contraction is ill-defined in the 4D spacetime; it is
synchronization dependent and consequently it is not an intrinsic
relativistic effect. The LT have nothing in common with the Lorentz
contraction; the LT cannot connect two \emph{spatial lengths} that are
simultaneously determined for relatively moving inertial observers. The
Lorentz contracted length and the rest length are two different quantities
and they are not related by the LT. Rohrlich [6] named such transformations
(Lorentz contraction) that do not refer to the same quantity - the
\textquotedblleft apparent\textquotedblright\ transformations, whereas the
transformations which refer to the same 4D quantity as the \textquotedblleft
true\textquotedblright\ transformations, e.g., the LT. It is visible from (%
\ref{ee}), (\ref{ut}) and (\ref{fr}) that the transformations of the
components of $\mathbf{E}$ and $\mathbf{B}$ do not refer to the same
quantity and therefore they are also the AT and not the true
transformations, i.e., the LT.

In [3] and [5] instead of the Lorentz contraction and the time dilation the
4D geometric quantities (GQs) are used, the position 4-vector, the distance
4-vector between two events and the spacetime length. In [5] it is shown
that all well-known experiments that test special relativity, e.g., the
\textquotedblleft muon\textquotedblright\ experiment, the Michelson-Morley
type experiments, the Kennedy-Thorndike type experiments and the
Ives-Stilwell type experiments are in a complete agreement, independently of
the chosen synchronization, with the 4D geometric approach, whereas it is
not the case with Einstein's approach with the Lorentz contraction and the
time dilation if the \textquotedblleft r\textquotedblright\ synchronization
is used.

In this paper, in section 2, the geometric algebra formalism, the standard
basis and the\textbf{\ }$\{r_{\mu }\}$ basis with the \textquotedblleft
r\textquotedblright\ synchronization are briefly discussed. In section 2.1,
some additional objections to the derivations of the AT are presented. In
sections 3.1 and 3.3 it is proved in a mathematically correct way that in
the 4D spacetime the electric and magnetic fields are not the usual 3D
fields $\mathbf{E}$ and $\mathbf{B}$ but that they are properly defined
vectors on the 4D spacetime, the 4D vectors $E$ and $B$. In the whole text $%
E $, $B$ will be simply called -\ vectors - or the 4D vectors, whereas the
usual $\mathbf{E}$, $\mathbf{B}$ will be called the 3D vectors. In sections
4.1 and 4.2 the proofs are given that the AT of the 3D fields are not the
mathematically correct LT, because the LT are properly defined on the 4D
spacetime and cannot transform the 3D quantities. The LT transform the
electric field vector in the same way as any other vector transforms, i.e.,
again to the electric field vector. Sections 3.1, 3.3, 4.1 and 4.2 are the
central sections and they contain the most important results that are
obtained in this paper. In sections 5.1 and 5.2, for the reader's
convenience, the derivations of the AT and the LT are compared using
matrices. In section 6, the derivation of the AT from the textbook by
Blandford and Thorne (BT) [7] is discussed and objected. In [7], in contrast
to, e.g., [1, 4], a geometric viewpoint is adopted; \emph{the physical laws
are stated as geometric, coordinate-free relationships between the
geometric, coordinate-free quantities. }Particularly, in section 1.10 in
[7], \emph{it is discussed the nature of electric and magnetic fields and
they are considered to be the 4D fields.} But, nevertheless, BT also derived
the AT of the 3D vectors $\mathbf{E}$ and $\mathbf{B}$, their equation
(1.113), and not the correct LT of the 4D fields, equations (\ref{aec}), (%
\ref{LTE}) and (\ref{el}) here. They have \emph{not }noticed that under the
LT the electric field 4D vector must transform as any other 4D vector
transforms. In section 7.1, it is discussed the derivation of the AT from
the paper by Klajn and Smoli\'{c} (KS) [8]. KS [8] use the tensor formalism
with the abstract index notation, but in section 3 in [8] they made almost
the same mistakes as in BT [7]. In section 7.2 similar shortcomings in the
treatment of the angular momentums and spin that are made in section 4 in
[8] are discussed and objected. In section 8, the mathematically correct
definitions with the 4D GQs of the orbital angular momentums and spins are
discussed. In section 9, the electromagnetic field of a point charge in
uniform motion is investigated and it is explicitly shown that 1) the
primary quantity is the bivector $F$\ (equations (\ref{fcf}) and (\ref{fmn}%
)) and 2) that the observer dependent 4D vectors $E$ and $B$, equation (\ref%
{ecb}), correctly describe both the electric and magnetic fields for all
relatively moving inertial observers and for all bases chosen by them. In
section 10, a brief discussion is presented of the comparison with the
experiments on the motional emf. It is shown that the theory with the 4D
quantities and their LT, equations (\ref{aec}), (\ref{LTE}) and (\ref{el})
here, is in agreement with the principle of relativity, equations (\ref{emf4}%
) and (\ref{ef5}), whereas it is not the case with the usual approach with
the 3D quantities and their AT, equations (\ref{eps}) - (\ref{eec}). In
section 11, the discussion of the obtained results is presented and the
conclusions are given.\bigskip \bigskip

\noindent \textbf{2. The geometric algebra formalism. The }$\{r_{\mu }\}$
\textbf{basis with the \textquotedblleft r\textquotedblright\ synchronization%
}\bigskip

\noindent Here, we shall also deal either with the abstract, coordinate-free
4D GQs, or with their representations in some basis, the 4D coordinate-based
geometric quantities (CBGQs) comprising \emph{both} components and \emph{a
basis}, e.g., the position vector, $x=x^{\nu }\gamma _{\nu }$. The
coordinate-free 4D GQs will be called the abstract quantities (AQs). \emph{%
An independent physical reality is attributed to the 4D GQs and not, as
usual, to the 3D quantities.} Every 4D CBGQ is \emph{invariant} under the
passive LT. \emph{The invariance of a 4D CBGQ under the passive LT reflects
the fact that such 4D GQ represents the same physical quantity for
relatively moving inertial observers.} We shall use the geometric algebra
formalism. The geometric (Clifford) product of two multivectors $A$ and $B$
is written by simply juxtaposing multivectors $AB$. For vectors $a$ and $b$
the geometric product $ab$\ decomposes as $ab=a\cdot b+a\wedge b$, where the
inner product $a\cdot b$\ is $a\cdot b\equiv (1/2)(ab+ba)$ and the outer (or
exterior) product $a\wedge b\ $is\ $a\wedge b\equiv (1/2)(ab-ba)$.\ For the
reader's convenience, all equations will be written with the CBGQs in the
standard basis. Therefore, the knowledge of the geometric algebra is not
required for the understanding of this presentation. The standard basis $%
\left\{ \gamma _{\mu }\right\} $ is a right-handed orthonormal frame of
vectors in the Minkowski spacetime $M^{4}$ with $\gamma _{0}$ in the forward
light cone, $\gamma _{0}^{2}=1$ and $\gamma _{k}^{2}=-1$ ($k=1,2,3$). The $%
\gamma _{\mu }$ generate by multiplication a complete basis for the
spacetime algebra: $1$, $\gamma _{\mu }$, $\gamma _{\mu }\wedge \gamma _{\nu
}$, $\gamma _{\mu }\gamma _{5}$, $\gamma _{5}$ ($2^{4}=16$ independent
elements). $\gamma _{5}$ is the right-handed unit pseudoscalar, $\gamma
_{5}=\gamma _{0}\wedge \gamma _{1}\wedge \gamma _{2}\wedge \gamma _{3}$. Any
multivector can be expressed as a linear combination of these 16 basis
elements of the spacetime algebra. The $\left\{ \gamma _{\mu }\right\} $
basis corresponds to Einstein's system of coordinates in which the Einstein
synchronization of distant clocks [9] and Cartesian space coordinates $x^{i}$
are used in the chosen inertial frame of reference. Here, we shall also
introduce another basis, the $\{r_{\mu }\}$ basis with the \textquotedblleft
r\textquotedblright\ synchronization. The \textquotedblleft
r\textquotedblright\ synchronization is commonly used in everyday life. If
the observers who are at different distances from the studio clock set their
clocks by the announcement from the studio\ then they have synchronized
their clocks with the studio clock according to the \textquotedblleft
r\textquotedblright\ synchronization.

The unit vectors in the $\{\gamma _{\mu }\}$ basis and the $\{r_{\mu }\}$
basis are connected as $r_{0}=\gamma _{0}$, $r_{i}=\gamma _{0}+\gamma _{i}$.
Hence, the components $g_{\mu \nu ,r}$ of the metric tensor are $g_{ii,r}=0$%
, and all other components are $=1$. Obviously it is completely different
than in the $\left\{ \gamma _{\mu }\right\} $ basis, i.e. than the Minkowski
metric, which, here, is chosen to be $g_{\mu \nu }=diag(1,-1,-1,-1)$. (Note
that in [3] and [5] the Minkowski metric is $g_{\mu \nu }=diag(-1,1,1,1)$.)
Then, according to equation (4) from [3], one can use $g_{\mu \nu ,r}$ to
find the transformation matrix $R_{\;\nu }^{\mu }$ that connects the
components in the $\left\{ \gamma _{\mu }\right\} $ and the $\{r_{\mu }\}$
bases. The only components that are different from zero are%
\begin{equation}
R_{\;\mu }^{\mu }=-R_{\;i}^{0}=1.  \label{er}
\end{equation}%
The inverse matrix $(R_{\;\nu }^{\mu })^{-1}$ connects the \textquotedblleft
old\textquotedblright\ basis, $\left\{ \gamma _{\mu }\right\} $, with the
\textquotedblleft new\textquotedblright\ one, $\{r_{\mu }\}$. The components
of any vector are connected in the same way as the components of the
position vector $x$ are connected, i.e., as%
\begin{equation}
x_{r}^{0}=x^{0}-x^{1}-x^{2}-x^{3},\quad x_{r}^{i}=x^{i}.  \label{ptr}
\end{equation}%
This reveals that \emph{in the} $\{r_{\mu }\}$\emph{\ basis the space} $%
\mathbf{r}$ \emph{and the time} $t$\ \emph{cannot be separated}; the
\textquotedblleft 3+1 split\textquotedblright\ of the spacetime into space +
time is \emph{impossible}. Note that there is the zeroth component of $x$\
in the $\{r_{\mu }\}$ basis, $x_{r}^{0}\neq 0$, even if in the standard
basis $x^{0}=0$, but the spatial components $x^{i}\neq 0$. This means that
\emph{in the 4D spacetime only the position vector} $x$, $x=x^{\mu }\gamma
_{\mu }=x_{r}^{\mu }r_{\mu }$, \emph{is properly defined quantity}. In
general, the position in the 3D space $\mathbf{r}$ and the time $t$\ have
not an independent reality in the 4D spacetime. Although the Einstein and
the \textquotedblleft r\textquotedblright\ synchronizations are completely
different they are equally well physical and relativistically correct
synchronizations. \emph{Every synchronization is only a convention and
physics must not depend on conventions.} An important consequence of the
result that in the 4D spacetime $\mathbf{r}$ and $t$\ are not well-defined
is presented in section 4 in [10]. There, it is shown that only the world
parity $W$, $Wx=-x$, is well defined in the 4D spacetime and not the usual $%
T $ and $P$ inversions. We remark that in order to treat different bases on
an equal footing the general transformation matrix $T_{\;\nu }^{\mu }$ is
presented in [3], equation (4), that connects the $\{\gamma _{\mu }\}$ basis
and some other basis, e.g., the $\{r_{\mu }\}$ basis, in the same reference
frame. That matrix $T_{\;\nu }^{\mu }$\ is expressed in terms of the basis
components of the metric tensor and for the connection with the $\{r_{\mu
}\} $ basis it is given by equation (\ref{er}). It is worth mentioning that
in equation (1) in [3] it is derived such form of the LT, which is
independent of the chosen system of coordinates, including different
synchronizations.\bigskip \bigskip

\noindent \textit{2.1. Other objections to the derivations of the AT\bigskip
}

\noindent 3) As already mentioned above (the objection 1)) the
identification of the components of $\mathbf{E}$ and $\mathbf{B}$ with the
components of $F^{\alpha \beta }$, (\ref{ieb}), is synchronization
dependent. If the components $F^{\alpha \beta }$ of $F$ are transformed by
the transformation matrix $R_{\;\nu }^{\mu }$ to the $\{r_{\mu }\}$ basis,
then it is obtained that, e.g.,%
\begin{equation}
F_{r}^{10}=F^{10}-F^{12}-F^{13}.  \label{are}
\end{equation}%
Hence, as shown in [3], [10], [2], in the $\left\{ r_{\mu }\right\} $ basis
the identification $E_{1r}=F_{r}^{10}$, as in (\ref{ieb}), yields that the
component $E_{1r}$ is expressed as the combination of $E_{i}$ and $B_{i}$
components from the $\left\{ \gamma _{\mu }\right\} $ basis

\begin{equation}
E_{1r}=F_{r}^{10},\quad E_{1r}=E_{1}+cB_{3}-cB_{2}.  \label{FEr}
\end{equation}%
This means that\textit{\ if the \textquotedblleft r\textquotedblright\
synchronization is used then it is not possible to make the usual
identifications} (\ref{ieb}) \textit{and} (\ref{eb2}).

4) As discussed in the next section, in the 4D geometric approach the
primary quantity for the whole electromagnetism is a physically measurable
quantity, the bivector field $F=(1/2)F^{\mu \nu }\gamma _{\mu }\wedge \gamma
_{\nu }$, where $\gamma _{\mu }\wedge \gamma _{\nu }$ is the bivector basis
and the basis components $F^{\mu \nu }$\ are determined as $F^{\mu \nu
}=\gamma ^{\nu }\cdot (\gamma ^{\mu }\cdot F)=(\gamma ^{\nu }\wedge \gamma
^{\mu })\cdot F$. In the same way as for any other CBGQ it holds that
bivector $F$ \emph{is the same 4D quantity} for relatively moving inertial
observers and for all bases chosen by them, e.g.,

\begin{equation}
F=(1/2)F^{\mu \nu }\gamma _{\mu }\wedge \gamma _{\nu }=(1/2)F_{r}^{\mu \nu
}r_{\mu }\wedge r_{\nu }=(1/2)F^{\prime \mu \nu }\gamma _{\mu }^{\prime
}\wedge \gamma _{\nu }^{\prime }=(1/2)F_{r}^{\prime \mu \nu }r_{\mu
}^{\prime }\wedge r_{\nu }^{\prime },  \label{fc}
\end{equation}%
where the primed quantities in both bases $\{\gamma _{\mu }\}$ and $\{r_{\mu
}\}$ are the Lorentz transforms of the unprimed ones. For the $\{r_{\mu }\}$
basis and the LT in that basis see [3]. \emph{Only the whole} $F$\ \emph{from%
} (\ref{fc}) \emph{is a mathematically correctly defined quantity} and it
does have a definite physical reality. The components $F^{i0}$, or $F^{ij}$
(implicitly determined in the standard basis $\{\gamma _{\mu }\}$), if taken
alone, are not properly defined physical quantities in the 4D spacetime. The
transformations of these components, e.g., equation (\ref{fe}), which are
extracted from the LT\ of the whole properly defined physical quantity $%
F=(1/2)F^{\alpha \beta }\gamma _{\alpha }\wedge \gamma _{\beta }$, are not
the relativistically correct LT and actually they have nothing to do with
the LT. They do not refer to \emph{the same }4\emph{D} \emph{quantity} for
relatively moving observers. Hence, the determination of $\mathbf{E}$ and $%
\mathbf{B}$ by the components $F^{i0}$ and $F^{ij}$, respectively, as the
quantities that do not depend on the 4-velocity of the observer is not
mathematically and relativistically correct. In contrast to it, the
determination of vectors $E$ and $B$ relative to the observer by the
decomposition of $F$, i.e., by equations (\ref{E2}) and (\ref{E1}) with
coordinate-free quantities, or (\ref{fm}) and (\ref{ebv}) with the CBGQs is
mathematically and relativistically correct. Every antisymmetric tensor of
the second rank (as a geometric quantity) can be decomposed into two vectors
and a unit timelike vector, in this case, $v/c$. This proves in another way
that the usual identification of the \emph{components} of $\mathbf{E}$ and $%
\mathbf{B}$ with the \emph{components} of $F^{\alpha \beta }$, (\ref{ieb}),
cannot have a definite physical sense; the components are coordinate
quantities and they are only a part of the representation in some basis of
an abstract, coordinate-free bivector $F$.

5) In addition, it is worth mentioning that in the usual covariant
approaches, e.g., [1], the components $F^{\alpha \beta }$ are defined in
terms of a 4-vector potential $A^{\alpha }=(\Phi ,\mathbf{A})$, equation
(11.132) in [1], as $F^{\alpha \beta }=\partial ^{\alpha }A^{\beta
}-\partial ^{\beta }A^{\alpha }$, equation (11.136) in [1]. The 3D fields $%
\mathbf{E}$ and $\mathbf{B}$ are determined in terms of the potentials by
equation (11.134) in [1], which, together with equation (11.136) in [1],
leads to equation (11.137) in [1] in which, as already stated, the
components $F^{\alpha \beta }$ are expressed in terms of the components of
the 3D vectors $\mathbf{E}$ and $\mathbf{B}$. According to that procedure
from [1]\ the 4-vector potential $A^{\alpha }$ (gauge dependent and thus
unmeasurable quantity) is considered to be the primary quantity which
determines the measurable quantities, the electric and magnetic fields and
also $F^{\alpha \beta }$. Observe that, contrary to the assertions from [1],
$A^{\alpha }$ is not a 4D vector. $A^{\alpha }$ are only components
implicitly taken in the standard basis of the 4D vector $A=A^{\mu }\gamma
_{\mu }$. In the 4D spacetime only the whole 4D potential $A=A^{\mu }\gamma
_{\mu }=A_{r}^{\mu }r_{\mu }$ is a well-defined quantity,\ whereas it is not
the case with the usual scalar potential $\Phi $\ and the 3D vector
potential $\mathbf{A}$ in which the components $A_{x,y,z}$ are multiplied by
the unit 3D vectors $\mathbf{i}$, $\mathbf{j}$, $\mathbf{k}$ and not by the
properly defined unit 4D vectors $\gamma _{\mu }$.\bigskip \bigskip

\noindent \textbf{3. The proofs that the electric and magnetic fields are
properly}

\textbf{defined vectors on the 4D spacetime and not the usual 3D fields}%
\bigskip \medskip

\noindent \textit{3.1.} \textit{Oziewicz's proof}\bigskip

\noindent There is a simple but very strong and completely correct
mathematical argument, which is stated by Oziewicz, e.g., in [11]:

\emph{What is essential for the number of components of a vector field is
the number of variables on which that vector field depends, i.e.,} \emph{the
dimension of its domain}.\emph{\ In general, the dimension of a vector field
that is defined on a n-dimensional space is equal - n. The electric and
magnetic fields are defined on a 4D space, i.e., the spacetime. They are
always functions of the position vector} $x$. \emph{This means that they are
not the usual 3D fields,\ but they are properly defined vectors on the 4D
spacetime}, $E(x)$ and $B(x)$. In any basis they have four components some
of which can be zero. This is a fundamental argument and it cannot be
disputed in any way. It is very surprising that this argument is not applied
in physics much earlier.

The mentioned argument holds in the same measure for the polarization vector
$P(x)$ and the magnetization vector $M(x)$, which are discussed in detail in
[12, 13, 2]. In [12] the electromagnetic field equations for moving media
are presented, whereas in [13] the constitutive relations and the
magnetoelectric effect for moving media are investigated from the geometric
point of view. $P(x)$ and $M(x)$ are also properly defined vectors on the 4D
spacetime and not the 3D vectors as usually considered, e.g., in [1, 4].
Note that in the 4D spacetime we \emph{always} have to deal with correctly
defined vectors $E(x)$, $B(x)$, $P(x)$, $M(x)$, etc. even in the usual
static case, i.e., if the usual 3D fields $\mathbf{E(\mathbf{r})}$, $\mathbf{%
B(r)}$ do not explicitly depend on the time $t$. The reason is that if \emph{%
in the 4D spacetime }the standard basis is used then\emph{\ }the LT cannot
transform the spatial coordinates from one frame only to spatial coordinates
in a relatively moving inertial frame of reference. What is static case for
one inertial observer is not more static case for relatively moving inertial
observer, but a time dependent case. Furthermore, if an observer uses the
\textquotedblleft r\textquotedblright\ synchronization and not the standard
Einstein's synchronization, then, as seen from (\ref{ptr}), the space and
time are not separated and the usual 3D vector $\mathbf{\mathbf{r}}$ is
meaningless. If the principle of relativity has to be satisfied and the
physics must be the same for all inertial observers and for $\{\gamma _{\mu
}\}$, $\{r_{\mu }\}$, $\{\gamma _{\mu }^{\prime }\}$, etc. bases which they
use, then the properly defined quantity is the position vector $x$,
\begin{equation}
x=x^{\nu }\gamma _{\nu }=x^{\prime \nu }\gamma _{\nu }^{\prime }=x_{r}^{\nu
}r_{\nu }=x_{r}^{\prime \nu }r_{\nu }^{\prime },  \label{iks}
\end{equation}%
and not $\mathbf{\mathbf{r}}$ and $t$. Consequently, in the 4D spacetime,
e.g., the electric field is properly defined as the vector $E(x)$ for which
the relation (\ref{erc}) given below holds.\bigskip \medskip

\noindent \textit{3.2.} \textit{Briefly about} \textit{the} $F$\ \textit{%
formulation}\bigskip

\noindent In [14] an axiomatic geometric formulation of electromagnetism
with only one axiom, the field equation for the bivector field $F$, equation
(4) in [14], is constructed. There, it is shown that the bivector $F=F(x)$,
which represent the electromagnetic field, can be taken as the primary
quantity for the whole electromagnetism. It yields a complete description of
the electromagnetic field and, in fact, there is no need to introduce either
the field vectors or the potentials. If the field equation for $F$\ is
written with AQs it becomes
\begin{equation}
\partial \cdot F+\partial \wedge F=j/\varepsilon _{0}c,  \label{fef}
\end{equation}%
where the source of the field is the charge-current density vector $j(x)$
(equation (4) in [14]). If $j(x)$\ is the sole source of $F$\ then the
general solution for $F$\ with \emph{AQs }is given by equation (8) in [14].
Particularly, the general expression for $F$ for an arbitrary motion of a
charge is given by equation (10) in [14] with AQs and as a CBGQ in the $%
\{\gamma _{\mu }\}$ basis by equation (11) in [14]. $F$\ of point charge in
uniform motion as an AQ is given by equation (12) in [14],\ i.e., equation (%
\ref{fcf}) here. The components in the standard basis $F^{\alpha \beta }$
from that equation (11) in [14] are the same as the usual result from
Chapter 14 in [1]. If the equation for $F$\ (\ref{fef}) is written with
CBGQs in the $\left\{ \gamma _{\mu }\right\} $ basis it becomes equation (5)
in [14],%
\begin{equation}
\partial _{\alpha }F^{\alpha \beta }\gamma _{\beta }-\partial _{\alpha }\
^{\ast }F^{\alpha \beta }\gamma _{5}\gamma _{\beta }=(1/\varepsilon
_{0}c)j^{\beta }\gamma _{\beta },  \label{mf}
\end{equation}%
where the usual dual tensor (components) is $^{\ast }F^{\alpha \beta
}=(1/2)\varepsilon ^{\alpha \beta \gamma \delta }F_{\gamma \delta }$. From
that equation one easily finds the usual covariant form (only the basis
components of the 4D geometric quantities in the $\left\{ \gamma _{\mu
}\right\} $ basis) of the field equations as equation (6) in [14],
\begin{equation}
\partial _{\alpha }F^{a\beta }=j^{\beta }/\varepsilon _{0}c,\quad \partial
_{\alpha }\ ^{\ast }F^{\alpha \beta }=0.  \label{mfc}
\end{equation}%
These two equations for \emph{the components in the standard basis}$\
F^{\alpha \beta }$\ are the equations (11.141) and (11.142) in [1].

In the same paper, [14], it is also shown that this formulation with the $F$%
\ field is in a complete agreement with the Trouton-Noble experiment, i.e.,
in the approach with $F$\ as a 4D GQ there is no Trouton-Noble paradox. It
is clearly visible from [14] and this short presentation that, in principle,
\emph{the components }$F^{\alpha \beta }$\ of the electromagnetic field
tensor, i.e., of the bivector $F$\ here and in [14], \emph{have nothing to
do with the components of the 3D vectors} $\mathbf{E}$ \emph{and} $\mathbf{B}
$. \emph{Only the whole} $F$\ \emph{has an independent physical reality; it
is a physically measurable quantity by the Lorentz force density}, $%
K_{(j)}=F\cdot j/c$, equation (27) in [14],\emph{\ or, for a charge} $q$
\emph{by the Lorentz force}

\begin{equation}
K_{L}=(q/c)F\cdot u,  \label{LF}
\end{equation}%
where $u$ is the 4D velocity vector of a charge $q$ (it is defined to be the
tangent to its world line).

It is worth noting that the expression for the Lorentz force density, $%
K_{(j)}=F\cdot j/c$, is directly derived from the field equation for $F$ (%
\ref{fef}). Similarly, in [14], the coordinate-free expressions for the
stress-energy vector $T(n)$ (equations (37) and (38)), the energy density $U$
(scalar, equation (39)), the Poynting vector $S$ (equation (40)), the
momentum density vector $g$ (equation (42)), the angular momentum density $M$
(bivector, equation (43)), the local charge conservation law (equation (48))
and the local energy-momentum conservation law (equations (49) and (50)) are
all directly derived from that field equation (\ref{fef}). In that axiomatic
geometric formulation from [14] $T(n)$\ is the most important quantity for
the momentum and energy of the electromagnetic field,%
\begin{equation}
T(n)=-(\varepsilon _{0}/2)\left[ (F\cdot F)n+2(F\cdot n)\cdot F\right] ,
\label{tns}
\end{equation}%
equation (37) in [14]. $T(n)$ is a vector-valued linear function on the
tangent space at each spacetime point $x$ describing the flow of
energy-momentum through a hypersurface with normal $n=n(x)$. It can be
expressed by $U$ and $S$ as in equation (41) in [14],%
\begin{eqnarray}
T(n) &=&Un+(1/c)S,\quad U=-(\varepsilon _{0}/2)\left[ (F\cdot F)+2(F\cdot
n)^{2}\right] ,  \notag \\
S &=&-\varepsilon _{0}c\left[ (F\cdot n)\cdot F-(F\cdot n)^{2}n\right]
\label{ust}
\end{eqnarray}%
Observe that $T(n)$ as a whole quantity, i.e., the combination of $U$ and $S$
from (\ref{ust}) enters into a fundamental physical law, the local
energy-momentum conservation law

\begin{equation}
\partial \cdot T(n)=0  \label{coti}
\end{equation}%
for the free fields, equation (49) in [14]. This means, as stated in [14],
that only $T(n)$, as a whole quantity, does have a physically correct
interpretation. In [14] this viewpoint is nicely illustrated considering an
apparent paradox in the usual 3D formulation in which the Poynting vector $S$
is interpreted as an energy flux due to the propagation of fields. If such
an interpretation of $S$\ is adopted then there is a paradox for the case of
an uniformly accelerated charge, e.g., section 6.8 in [1]. In that case, $%
S=0 $ (there is no energy flow) but at the same time $U\neq 0$ (there is an
energy density) for the field points on the axis of motion. The obvious
question is how the fields propagate along the axis of motion to give that $%
U\neq 0$. In the formulation with 4D GQs the important quantity is $T(n)$
and not $S$ and $U$ taken separately. $T(n)$ is $\neq 0$ everywhere on the
axis of motion and the local energy-momentum conservation law (\ref{coti})
holds everywhere.\bigskip \medskip

\noindent \textit{3.3.} \textit{Proof by the use of the decomposition of} $F$%
\bigskip

\noindent In contrast to the usual covariant approach, which deals with the
identification of components (\ref{ieb}) and (\ref{eb2}), it is possible to
construct in a mathematically correct way the 4D vectors of the electric and
magnetic fields using the decomposition of $F$. There is \emph{a
mathematical theorem according to which any antisymmetric tensor of the
second rank can be decomposed into two space-like vectors and the unit
time-like vector.} For the proof of that theorem in geometric terms see,
e.g., [15].

If that theorem is applied to the bivector $F$ then it is obtained that

\begin{equation}
F=E\wedge v/c+(IcB)\cdot v/c,  \label{E2}
\end{equation}%
where the electric and magnetic fields are represented by vectors $E(x)$ and
$B(x)$, see, e.g., [14]. The unit pseudoscalar $I$ is defined algebraically
without introducing any reference frame. If $I$ is represented in the $%
\left\{ \gamma _{\mu }\right\} $ basis it becomes $I=\gamma _{0}\wedge
\gamma _{1}\wedge \gamma _{2}\wedge \gamma _{3}=\gamma _{5}$. The vector $v$
in the decomposition (\ref{E2}) is interpreted as the velocity vector of the
observers who measure $E$ and $B$ fields. Then $E(x)$ and $B(x)$ are defined
with respect to $v$, i.e., with respect to the observer, as
\begin{equation}
E=F\cdot v/c,\quad B=-(1/c)I(F\wedge v/c).  \label{E1}
\end{equation}%
It also holds that $E\cdot v=B\cdot v=0$; both $E$ and $B$ are space-like
vectors. If the decomposition (\ref{E2}) is written with the CBGQs in the $%
\{\gamma _{\mu }\}$ basis it becomes

\begin{equation}
F=(1/2)F^{\mu \nu }\gamma _{\mu }\wedge \gamma _{\nu },\ F^{\mu \nu
}=(1/c)(E^{\mu }v^{\nu }-E^{\nu }v^{\mu })+\varepsilon ^{\mu \nu \alpha
\beta }v_{\alpha }B_{\beta },  \label{fm}
\end{equation}%
where $\gamma _{\mu }\wedge \gamma _{\nu }$ is the bivector basis. If the
equations for $E$ and $B$ (\ref{E1}) are written with the CBGQs in the $%
\{\gamma _{\mu }\}$ basis they become%
\begin{equation}
E=E^{\mu }\gamma _{\mu }=(1/c)F^{\mu \nu }v_{\nu }\gamma _{\mu },\quad
B=B^{\mu }\gamma _{\mu }=(1/2c^{2})\varepsilon ^{\mu \nu \alpha \beta
}F_{\nu \alpha }v_{\beta }\gamma _{\mu }.  \label{ebv}
\end{equation}%
All these relations, (\ref{E2}) - (\ref{ebv}) are the mathematically correct
definitions. They are first reported (only components implicitly taken in
the standard basis) by Minkowski in section 11.6 in [16].

Let us introduce the $\gamma _{0}$ - frame; the frame of \textquotedblleft
fiducial\textquotedblright\ observers for which $v=c\gamma _{0}$ and in
which the standard basis is chosen. Therefore, in the $\gamma _{0}$-frame,
e.g., $E$ becomes $E=F\cdot \gamma _{0}$. It can be shown that in the $%
\gamma _{0}$ - frame $E\cdot \gamma _{0}=B\cdot \gamma _{0}=0$, which means
that $E$ and $B$ are orthogonal to $\gamma _{0}$; they refer to the 3D
subspace orthogonal to the specific timelike direction $\gamma _{0}$. If $E$
and $B$ are written as CBGQs in the standard basis they become%
\begin{eqnarray}
E &=&E^{\mu }\gamma _{\mu }=0\gamma _{0}+F^{i0}\gamma _{i},  \notag \\
B &=&B^{\mu }\gamma _{\mu }=0\gamma _{0}+(1/2c)\varepsilon
^{0ijk}F_{kj}\gamma _{i}.  \label{gnl}
\end{eqnarray}%
Note that $\gamma _{0}=(\gamma _{0})^{\mu }\gamma _{\mu }$ with $(\gamma
_{0})^{\mu }=(1,0,0,0)$. Hence, in the $\gamma _{0}$-frame the temporal
components of $E$ and $B$ are zero and only the spatial components remain%
\begin{equation}
E^{0}=B^{0}=0,\quad E^{i}=F^{i0},\quad B^{i}=(1/2c)\varepsilon ^{0ijk}F_{kj}.
\label{sko}
\end{equation}%
It is visible from (\ref{gnl}) and (\ref{sko}) that $E^{i}$ and $B^{i}$ are
the same as the components of the 3D $\mathbf{E}$ and $\mathbf{B}$, equation
(\ref{ieb}), i.e., the same as in equation (11.137) in [1]. However, there
are very important differences between the identifications (\ref{ieb}) and
equations (\ref{gnl}) and (\ref{sko}). The components of $\mathbf{E}$ and $%
\mathbf{B}$ in (\ref{ieb}) are not the spatial components of the 4D
quantities. They transform according to the AT (\ref{ee}). The antisymmetric
$\varepsilon $ tensor in (\ref{ieb}) and (\ref{eb2}) is a third-rank
antisymmetric tensor. On the other hand, the components of $E$ and $B$ in (%
\ref{gnl}) and (\ref{sko}) are the spatial components of the 4D geometric
quantities that are taken in the standard basis. They transform according to
the LT that are given below, equation (\ref{LTE}). The antisymmetric $%
\varepsilon $ tensor in (\ref{gnl}) and (\ref{sko}) is a fourth-rank
antisymmetric tensor. Furthermore, it is shown above, equations (\ref{are})
and (\ref{FEr}), that the identifications (\ref{ieb}) and (\ref{eb2}) do not
hold in the $\{r_{\mu }\}$ basis. But, the relations (\ref{ebv}) hold for
any chosen basis, including the $\{r_{\mu }\}$ basis, e.g.,
\begin{equation}
E=E^{\nu }\gamma _{\nu }=E_{r}^{\nu }r_{\nu }=(1/c)F_{r}^{\mu \nu }v_{\nu
,r}r_{\mu }.  \label{efr}
\end{equation}%
This can be easily checked using the above mentioned matrix $R_{\;\nu }^{\mu
}$.\ Thus, for the components of vector $E$ it also holds that
\begin{equation}
E_{r}^{0}=E^{0}-E^{1}-E^{2}-E^{3},\quad E_{r}^{i}=E^{i}.  \label{ei}
\end{equation}%
From these relations it follows that there is the zeroth component of $E$\
in the $\{r_{\mu }\}$ basis, $E_{r}^{0}\neq 0$, even if it is $=0$ in the
standard basis, $E^{0}=0$, but the spatial components $E^{i}\neq 0$. This
again shows that the components taken alone are not physical. The whole
consideration presented here explicitly reveals that in the 4D spacetime the
usual identifications (\ref{ieb}) and (\ref{eb2}) are not mathematically
correct and that

\emph{the electric field} $E$ \emph{is a vector (4D} \emph{vector); it is an
inner product of a bivector} $F$\ \emph{and the velocity vector} $v$ \emph{%
of the observer who measures fields.}

It is worth mentioning that in the 4D spacetime the mathematically correct
relations (\ref{E2}) - (\ref{ebv}) are already firmly theoretically founded
and they are known to many physicists. The recent example is in [17]; it is
only the electric part (the magnetic part is zero there). Similarly, in the
component form these relations are presented, e.g., in [18] and in the
basis-free form with the abstract 4D quantities in [7, 8, 15] and in, e.g.,
[19]. But, it has to be noted that from all of them only Oziewicz, see [11]
and references to his papers in it, \emph{exclusively} deals with the
abstract, basis-free 4D quantities. He correctly considers from the outset
that in the 4D spacetime such quantities are physical quantities and not the
usual 3D quantities. All others, starting with Minkowski [16], are not
consistent in the use of the 4D electric and magnetic fields. They use
together the 4D fields and the usual 3D fields $\mathbf{E}$ and $\mathbf{B}$
considering that the 3D fields are physically measurable quantities and that
their AT are the correct LT. Minkowski [16] introduced only in section 11.6
the 4D fields and their LT. In other sections he also dealt with the 3D
fields and their AT. \bigskip \bigskip

\noindent \textbf{4.\ The proofs that under the mathematically correct LT
the electric }

\textbf{field vector transforms as any other vector transforms, i.e., again }

\textbf{to the electric field vector\bigskip }

\noindent As proved in section 2 the electric field is properly defined
vector on the 4D spacetime and the same holds for the magnetic field. Hence,
\emph{under the LT},\emph{\ }e.g., \emph{the electric field vector must
transform as any other vector transforms}, i.e., \emph{again} \emph{to the
electric field vector}; \emph{there is no mixing with the magnetic field
vector }$B$. In [20] the same result is obtained for the electric field as a
bivector and for the magnetic field as well. This will be explicitly shown
both for the active LT in 4.1 and for the passive LT in 4.2.\bigskip \medskip

\noindent \textit{4.1.} \textit{Proof with the coordinate-free quantities,
AQs, and the active LT\bigskip }

\noindent Regarding the correct LT let us start from the definition with the
coordinate-free quantities $E=c^{-1}F\cdot v$ and with \emph{the active LT}.
\emph{Mathematically, }as noticed by Oziewicz [11], \emph{an active LT must
act on all tensor fields from which the vector field }$E$\emph{\ is
composed, including an observer's time-like vector field}.\ This means that
the mathematically correct active LT of $E=c^{-1}F\cdot v$\ are $E^{\prime
}=c^{-1}F^{\prime }\cdot v^{\prime }$; both $F$\ and $v$\ are transformed.
It was first discovered by Minkowski in section 11.6 in [16] but with
components implicitly taken in the standard basis and reinvented and
generalized in terms of 4D GQs in [21-26] and [20], see also section 5 in
[2]. As explicitly shown, e.g., in [26], in the geometric algebra formalism
\emph{any multivector} $N$ \emph{transforms by the active LT in the same way,%
} i.e., as $N\rightarrow N^{\prime }=RN\widetilde{R}$, where $R$\ is given
by equation (10) in [26] (equation (39) in [2]); for boosts in an arbitrary
direction the rotor $R$ is
\begin{equation}
R=(1+\gamma +\gamma \gamma _{0}\beta )/(2(1+\gamma ))^{1/2},  \label{LTR}
\end{equation}%
where $\gamma =(1-\beta ^{2})^{-1/2}$, the vector $\beta $ is $\beta =\beta
n $, $\beta $ on the r.h.s. of that equation is the scalar velocity in units
of $c$ and $n$ is not the basis vector but any unit space-like vector
orthogonal to $\gamma _{0}$. The reverse $\widetilde{R}$ is defined by the
operation of reversion according to which $\widetilde{AB}=\widetilde{B}%
\widetilde{A}$, for any multivectors $A$ and $B$, see section 3 in [26]
(section 5 in [2]). Hence, the vector $E=c^{-1}F\cdot v$\ transforms by
\emph{the mathematically correct} active LT $R$ into%
\begin{equation}
E^{\prime }=RE\widetilde{R}=c^{-1}R(F\cdot v)\widetilde{R}=c^{-1}(RF%
\widetilde{R})\cdot (Rv\widetilde{R})=c^{-1}F^{\prime }v^{\prime }.
\label{Lt}
\end{equation}%
If $v=c\gamma _{0}$ is taken in the expression for $E$\ then $E$ becomes $%
E=F\cdot \gamma _{0}$ and it transforms as in [16], i.e., that \emph{both} $%
F $ \emph{and} $\gamma _{0}$ are transformed by the LT.
\begin{equation}
E=F\cdot \gamma _{0}\longrightarrow E^{\prime }=R(F\cdot \gamma _{0})%
\widetilde{R}=(RF\widetilde{R})\cdot (R\gamma _{0}\widetilde{R}).  \label{Lg}
\end{equation}%
Hence, the explicit form for $E^{\prime }$\ with the abstract,
coordinate-free quantities is given by equation (13) in [26],
\begin{equation}
E^{\prime }=E+\gamma (E\cdot \beta )\{\gamma _{0}-(\gamma /(1+\gamma ))\beta
\}.  \label{aec}
\end{equation}%
In (\ref{aec}) $\beta $ is a vector. In the standard basis and for boosts in
the direction $x^{1}$ the components of that $E^{\prime }$\ are
\begin{equation}
E^{\prime \mu }=(E^{\prime 0}=-\beta \gamma E^{1},\ E^{\prime 1}=\gamma
E^{1},\ E^{\prime 2,3}=E^{2,3}).  \label{LTE}
\end{equation}%
Under the active LT the electric field vector $E=F\cdot \gamma _{0}$ (as a
CBGQ $E=E^{\mu }\gamma _{\mu }=0\gamma _{0}+F^{i0}\gamma _{i}$) is
transformed into a new electric field vector $E^{\prime }$, (\ref{aec}).
Note that under the active LT the components are changed, (\ref{LTE}), but
the basis remains unchanged,
\begin{equation}
E^{\prime \nu }\gamma _{\nu }=-\beta \gamma E^{1}\gamma _{0}+\gamma
E^{1}\gamma _{1}+E^{2}\gamma _{2}+E^{3}\gamma _{3},  \label{eb}
\end{equation}%
see equation (14) in [26] (equation (43) in [2]), i.e., equation (\ref{an})
below. \emph{The components} $E^{\mu }$ \emph{transform by the LT again to
the components} $E^{\prime \mu }$ \emph{and there is no mixing with }$B^{\mu
}$. In general, the LT of the components $E^{\mu }$ (in the $\{\gamma _{\mu
}\}$ basis) of $E=E^{\mu }\gamma _{\mu }$ are given as
\begin{equation}
E^{\prime 0}=\gamma (E^{0}-\beta E^{1}),\ E^{\prime 1}=\gamma (E^{1}-\beta
E^{0}),\ E^{\prime 2,3}=E^{2,3},  \label{el}
\end{equation}%
for a boost along the $x^{1}$ axis, i.e., the same LT as for any other 4D
vector.

On the other hand, if in $E=F\cdot \gamma _{0}$ only $F$ is transformed by
the active LT and not $\gamma _{0}$, which is not a mathematically correct
procedure, then the components of that $E_{F}^{\prime }$\ will be denoted as
$E_{F}^{\prime \mu }$\ and they are
\begin{equation}
E_{F}^{\prime \mu }=(E_{F}^{\prime 0}=0,E_{F}^{\prime 1}=E^{1},E_{F}^{\prime
2}=\gamma (E^{2}-c\beta B^{3}),E_{F}^{\prime 3}=\gamma (E^{3}+c\beta B^{2})),
\label{ut}
\end{equation}%
see equation (17) in [26] (equation (46) in [2]), i.e., (\ref{Em4}) below.
\emph{The transformations of the spatial components (taken in the standard
basis) of} $E$ \emph{are exactly the same as the transformations of }$%
E_{x,y,z}$ \emph{from equation }(11.148)\emph{\ in }[1], i.e., as in
equation (\ref{ee}). However, from $E=F\cdot \gamma _{0}$ it follows that
the components of $E$\ are $E^{\mu }=(E^{0}=0$, $E^{1}$, $E^{2}$, $E^{3})$.
Hence,\emph{\ if only} $F$ \emph{is transformed by the LT then the temporal
components of both} $E$ \emph{and} $E_{F}^{\prime }$ \emph{are zero}, $%
E^{0}=E_{F}^{\prime 0}=0$, which explicitly reveals that such
transformations are not the mathematically correct LT; the LT cannot
transform $E^{0}=0$ again to $E_{F}^{\prime 0}=0$. This proves that the
transformations (\ref{LTE}) in which \emph{both} $F$ \emph{and} $\gamma _{0}$
are transformed are the correct LT.\bigskip \medskip

\noindent \textit{4.2.} \textit{Proof with CBGQs and the passive LT\bigskip }

\noindent If $E$\ is written as a CBGQ, i.e., as in (\ref{ebv}),$\ $then we
have to use the passive LT. For example, in the $\gamma _{0}$-frame $E$\ is
given as
\begin{equation}
E=E^{\mu }\gamma _{\mu }=[(1/c)F^{i0}v_{0}]\gamma _{i}=0\gamma
_{0}+E^{i}\gamma _{i}  \label{te1}
\end{equation}%
For boosts in the $\gamma _{1}$\ direction and if \emph{both} $F^{i0}$\
\emph{and} $v_{0}$\ are transformed by the LT then, as for any other CBGQ,
it holds that
\begin{equation}
E=E^{\mu }\gamma _{\mu }=[(1/c)F^{\prime \mu \nu }v_{\nu }^{\prime }]\gamma
_{\mu }^{\prime }=E^{\prime \mu }\gamma _{\mu }^{\prime },  \label{tec}
\end{equation}%
where, again, the components $E^{\prime \mu }$\ are the same as in (\ref{LTE}%
), see [24]. On the other hand, if only $F^{i0}$ is transformed but not $%
v_{0}$\ the transformed components $E_{F}^{\prime \mu }$\ are again the same
as in (\ref{ee}) and the same objections as in section 4.1 hold also here.
In addition, it can be easily checked that%
\begin{equation}
E_{F}^{\prime \mu }\gamma _{\mu }^{\prime }\neq E^{\mu }\gamma _{\mu },
\label{fr}
\end{equation}%
which additionaly proves that the transformations in which only $F$ is
transformed are not the relativistically correct LT. In that way it is also
proved that the transformations given by equations (11.148) ((\ref{ut})
here) and (11.149) from [1] are not the LT but, as called here, the
mathematically incorrect AT that do not refer to the same quantity. As can
be seen from the above discussion if $E$\ is written as a CBGQ then, as for
any other 4D CBGQ, it holds that
\begin{equation}
E=E^{\nu }\gamma _{\nu }=E^{\prime \nu }\gamma _{\nu }^{\prime }=E_{r}^{\nu
}r_{\nu }=E_{r}^{\prime \nu }r_{\nu }^{\prime }.  \label{erc}
\end{equation}%
Here, as in (\ref{fc}), the primed quantities in both bases $\{\gamma _{\mu
}\}$ and $\{r_{\mu }\}$ are the Lorentz transforms of the unprimed
ones.\bigskip \medskip

\noindent \textit{4.3. A short discussion of the field equations with
vectors }$E$ \textit{and} $B$\textit{\bigskip }

\noindent If the decomposition of $F$\ from (\ref{fm}) is introduced into (%
\ref{mf}) then the field equation (\ref{maeb}) is obtained

\begin{align}
\lbrack \partial _{\alpha }(\delta _{\quad \mu \nu }^{\alpha \beta }E^{\mu
}v^{\nu }+\varepsilon ^{\alpha \beta \mu \nu }v_{\mu }cB_{\nu })-& (j^{\beta
}/\varepsilon _{0})]\gamma _{\beta }+  \notag \\
\partial _{\alpha }(\delta _{\quad \mu \nu }^{\alpha \beta }v^{\mu }cB^{\nu
}+\varepsilon ^{\alpha \beta \mu \nu }v_{\mu }E_{\nu })\gamma _{5}\gamma
_{\beta }& =0,  \label{maeb}
\end{align}%
where $E^{\alpha }$ and $B^{\alpha }$ are the basis components in the
standard basis of the 4D vectors $E$ and $B$, $\delta _{\quad \mu \nu
}^{\alpha \beta }=\delta _{\,\,\mu }^{\alpha }\delta _{\,\,\nu }^{\beta
}-\delta _{\,\,\nu }^{\alpha }\delta _{\,\mu }^{\beta }$ and $\gamma _{5}$
is the pseudoscalar in the $\{\gamma _{\mu }\}$ basis. This is equation (40)
in [23], but there it is written using some unspecified basis $\left\{
e_{\mu }\right\} $. The first part in (\ref{maeb}) comes from $\partial
\cdot F=j/\varepsilon _{0}c$ and the second one (the source-free part) comes
from $\partial \wedge F=0$. As discussed in detail in [23] \emph{equation} (%
\ref{maeb}) \emph{is the relativistically correct, manifestly covariant
field equation} that generalizes the usual Maxwell equations with the 3D
fields $\mathbf{E}$ and $\mathbf{B}$. It, (\ref{maeb}), can be compared with
the usual formulation with the 3D quantities going to the $\gamma _{0}$%
-frame in which $v=c\gamma _{0}$ and equation (\ref{sko}) holds. This yields
that equation (\ref{maeb}) becomes

\begin{align}
(\partial _{k}E^{k}-j^{0}/c\varepsilon _{0})\gamma _{0}+(-\partial
_{0}E^{i}+c\varepsilon ^{ijk0}\partial _{j}B_{k}-j^{i}/c\varepsilon
_{0})\gamma _{i}+&  \notag \\
(-c\partial _{k}B^{k})\gamma _{5}\gamma _{0}+(c\partial
_{0}B^{i}+\varepsilon ^{ijk0}\partial _{j}E_{k})\gamma _{5}\gamma _{i}& =0.
\label{MEC}
\end{align}%
The equation (\ref{MEC}) contains all four usual Maxwell equations in the
component form. The first part (with $\gamma _{\alpha }$) in (\ref{MEC})
contains \emph{two} \emph{Maxwell equations} in the \emph{component form},
the Gauss law for the electric field (the first bracket, with $\gamma _{0}$)
and the Amp\`{e}re-Maxwell law (the second bracket, with $\gamma _{i}$). The
second part (with $\gamma _{5}\gamma _{\alpha }$) contains the \emph{%
component form} of another \emph{two Maxwell equations}, the Gauss law for
the magnetic field (with $\gamma _{5}\gamma _{0}$) and Faraday's law (with $%
\gamma _{5}\gamma _{i}$).

Observe that the component form of the Maxwell equations with the 3D\textbf{%
\ }$\mathbf{E}$\textbf{\ }and $\mathbf{B}$
\begin{eqnarray}
\partial _{k}E_{k}-j^{0}/c\varepsilon _{0} &=&0,\quad -\partial
_{0}E_{i}+c\varepsilon _{ikj}\partial _{j}B_{k}-j^{i}/c\varepsilon _{0}=0,
\notag \\
\partial _{k}B_{k} &=&0,\quad c\partial _{0}B_{i}+\varepsilon _{ikj}\partial
_{j}E_{k}=0  \label{j3}
\end{eqnarray}%
is obtained from the covariant Maxwell equations (\ref{mfc}) using the usual
identifications of six independent components of $F^{\mu \nu }$ with three
components $E_{i}$ and three components $B_{i}$ as in (\ref{ieb}) and also
in (\ref{eb2}). But, as shown above, such an identification is meaningless
in the $\left\{ r_{\mu }\right\} $ basis, which means that Maxwell equations
(\ref{j3}) do not hold in the $\left\{ r_{\mu }\right\} $ basis. Moreover,
the components of the 3D fields from (\ref{j3}) transform according to the
AT (\ref{ee}) and not according to mathematically correct LT (\ref{aec}) - (%
\ref{el}), which causes, as explicitly shown in [23], that equations (\ref%
{j3}) are not covariant under the LT. On the other hand, contrary to the
formulation of the electromagnetism with $\mathbf{E}$ and $\mathbf{B}$,

\emph{the formulation with the 4D fields }$E$ \emph{and} $B$, \emph{i.e.,
with equation }(\ref{maeb}), \emph{is correct not only in the} $\gamma _{0}$
- \emph{frame with the standard basis} $\left\{ \gamma _{\mu }\right\} $
\emph{but in all other relatively moving frames and it holds for any
permissible choice of coordinates, i.e., bases}.

This consideration reveals that the 4D fields $E$ and $B$ that transform
like in (\ref{aec}) - (\ref{el}) and the field equation (\ref{maeb}) do not
have the same physical interpretation as the usual 3D fields $\mathbf{E}$
and $\mathbf{B}$ and the usual Maxwell equations (\ref{j3}) except in the $%
\gamma _{0}$ - frame with the $\left\{ \gamma _{\mu }\right\} $ basis in
which $E^{0}=B^{0}=0$.

Here, it is at place a remark about the $\gamma _{0}$ - frame. The
dependence of the relations (\ref{ebv}) and the field equation (\ref{maeb})
on $v$ reflects the arbitrariness in the selection of the $\gamma _{0}$ -
frame, but at the same time this arbitrariness makes that equations (\ref%
{ebv}) and (\ref{maeb}) are independent of that choice. The $\gamma _{0}$ -
frame can be selected at our disposal depending on the considered problem
which proves that we don't have a kind of \textquotedblleft
preferred\textquotedblright\ frame theory. Some examples will be discussed
in sections 9 and 10.\bigskip \medskip

\noindent \textit{4.4. The\ generalization of the field equation for} $F$ (%
\ref{fef}) \textit{to a magnetized and polarized moving medium\bigskip }

\noindent The\ generalization of the field equation for $F$ (\ref{fef}) to a
magnetized and polarized moving medium with the generalized
magnetization-polarization bivector $\mathcal{M(}x\mathcal{)}$ is presented
in [12]. That generalization is obtained simply replacing $F$ by $F+\mathcal{%
M}/\varepsilon _{0}$, which yields the primary equations for the
electromagnetism in moving media

\begin{equation}
\partial (\varepsilon _{0}F+\mathcal{M})=j^{(C)}/c;\quad \partial \cdot
(\varepsilon _{0}F+\mathcal{M})=j^{(C)}/c,\ \partial \wedge F=0,  \label{F41}
\end{equation}%
equation (7) in [12]. $j^{(C)}$ is the conduction current density of the
\emph{free} charges and $j^{(\mathcal{M})}=-c\partial \cdot \mathcal{M}$ is
the magnetization-polarization current density of the \emph{bound} charges.
The total current density vector $j$ is $j=j^{(C)}+j^{(\mathcal{M})}$. If
written with the CBGQs in the standard basis that equation becomes

\begin{equation}
\partial _{\alpha }(\varepsilon _{0}F^{\alpha \beta }+\mathcal{M}^{\alpha
\beta })\gamma _{\beta }-\partial _{\alpha }(\varepsilon _{0}\ ^{\ast
}F^{\alpha \beta })\gamma _{5}\gamma _{\beta }=c^{-1}j^{(C)\beta }\gamma
_{\beta },  \label{F8}
\end{equation}%
what is equation (8) in [12]. Observe that if in equation for $F$\ (\ref{fef}%
) $j=j^{(C)}+j^{(\mathcal{M})}$ is the total current density then (\ref{fef}%
), \textit{i.e.}, (\ref{mf}), \emph{holds unchanged in moving medium as well.%
}

In the same way as in (\ref{fm}) the generalized magnetization-polarization
bivector $\mathcal{M(}x\mathcal{)}$ can be decomposed into two vectors, the
polarization vector $P(x)$ and the magnetization vector $M(x)$ and the unit
time-like vector $u/c$, equation (21) in [12],

\begin{equation}
\mathcal{M}=P\wedge u/c+(MI)\cdot u/c^{2},  \label{M1}
\end{equation}%
or, with the CBGQs in the $\{\gamma _{\mu }\}$ basis, equation (22) in [12],%
\begin{equation}
\mathcal{M}=(1/2)\mathcal{M}^{\mu \nu }\gamma _{\mu }\wedge \gamma _{\nu },\
\mathcal{M}^{\mu \nu }=(1/c)(P^{\mu }u^{\nu }-P^{\nu }u^{\mu
})+(1/c^{2})\varepsilon ^{\mu \nu \alpha \beta }M_{\alpha }u_{\beta }.
\label{mp}
\end{equation}%
The vector $u$ is identified with bulk velocity vector of the medium in
spacetime. Hence, as in (\ref{ebv}), equation (24) in [12],%
\begin{equation}
P=(1/c)\mathcal{M}^{\mu \nu }u_{\nu }\gamma _{\mu },\quad M=(1/2)\varepsilon
^{\mu \nu \alpha \beta }\mathcal{M}_{\alpha \nu }u_{\beta }\gamma _{\mu },
\label{mp1}
\end{equation}%
with $P^{\mu }u_{\mu }=M^{\mu }u_{\mu }=0$, only three components of $P$ and
three components of $M$ are independent since $\mathcal{M}$ is
antisymmetric. Inserting the decompositions of $F(x)$ (\ref{fm}) and $%
\mathcal{M(}x\mathcal{)}$ (\ref{mp}) into the field equation (\ref{F8}) one
finds equation (29) in [12]

\begin{equation}
\partial _{\alpha }\{\varepsilon _{0}[\delta _{\quad \mu \nu }^{\alpha \beta
}E^{\mu }v^{\nu }+c\varepsilon ^{\alpha \beta \mu \nu }v_{\mu }B_{\nu
}]+[\delta _{\quad \mu \nu }^{\alpha \beta }P^{\mu }u^{\nu
}+(1/c)\varepsilon ^{\alpha \beta \mu \nu }M_{\mu }u_{\nu }]\}\gamma _{\beta
}=j^{(C)\beta }\gamma _{\beta },  \label{I1}
\end{equation}%
where $\delta _{\quad \mu \nu }^{\alpha \beta }=\delta _{\,\,\mu }^{\alpha
}\delta _{\,\,\nu }^{\beta }-\delta _{\,\,\nu }^{\alpha }\delta _{\,\,\mu
}^{\beta }$. This is the part of the equation (\ref{F8}) with sources,
whereas another part, the equation without sources, equation (30) in [12],
becomes

\begin{equation}
\partial _{\alpha }(c\delta _{\quad \mu \nu }^{\alpha \beta }B^{\mu }v^{\nu
}+\varepsilon ^{\alpha \beta \mu \nu }E_{\mu }v_{\nu })\gamma _{5}\gamma
_{\beta }=0.  \label{I2}
\end{equation}%
The eqations (\ref{I1}) and (\ref{I2}) are the fundamental equations for
moving media and \emph{they replace all usual Maxwell's equations (with 3D
vectors) for moving media}. As stated in [12], in contrast to all usual
formulations of the field equations for moving media, the equation (\ref{I1}%
) contains two different velocity vectors, $v$ - the velocity of the
observers and $u$ - the velocity of the moving medium, which come from the
decompositions of $F$ and $\mathcal{M}$, equations (\ref{fm}) and (\ref{mp}%
), respectively. It is shown in [12] that, in the same way as for vacuum,
the field equations (\ref{I1}) and (\ref{I2}) with the 4D fields are not
equivalent to the usual Maxwell's equations (with 3D vectors) for moving
media because the AT of the 3D fields are not the mathematically correct LT.

Furthermore, in the same way as for vacuum, i.e., as in [14], one can derive
from (\ref{F41}) the stress-energy vector $T(n)$ for a moving medium simply
replacing $F$ by $F+\mathcal{M}/\varepsilon _{0}$ in equations (26), (37-47)
in [14], i.e., in equations (\ref{tns}), (\ref{ust}) here. The expression
for $T(n)$, $T(n)=Un+(1/c)S$, will remain unchanged, but the energy density $%
U$ and the Poynting vector $S$ will change according to the described
replacement. This will be important in the discussion of Abraham-Minkowski
controversy.\bigskip \bigskip

\noindent \textbf{5. The comparison of the derivations of the AT and the LT
using}

\textbf{matrices (the components in the standard basis)}\bigskip \medskip

\noindent \textit{5.1. The electric and magnetic fields as vectors\bigskip }

\noindent For the reader's convenience the same results as in sections 3 -
3.3 can be obtained explicitly using the matrices. We write the relation $%
E^{\mu }=c^{-1}F^{\mu \nu }v_{\nu }$ in the $\gamma _{0}$ - frame, i.e., for
$v=c\gamma _{0}$. From the matrix for $F^{\mu \nu }$ and $v_{\nu }=(c,0,0,0)$
one finds $E^{\mu }=(0,F^{10}=E^{1},F^{20}=E^{2},F^{30}=E^{3})$.

Then, \emph{for the AT only} $F^{\mu \nu }$ \emph{is transformed by the LT
but not the velocity of the observer} $v=c\gamma _{0}$. The Lorentz
transformed $F^{\mu \nu }$ is (symbolically) $F^{\prime }=AF\widetilde{A}$;
here $A$, $F$, .. denote matrices. This relation can be written with
components as $F^{\prime \mu \nu }=A_{\rho }^{\mu }F^{\rho \sigma }%
\widetilde{A}_{\sigma }^{\nu }$. The matrix $A$\ is the boost in the
direction $x^{1}$ (in the standard basis) and it is written in equation (\ref%
{an}). $A$\ is also given by equation (11.98)\ in [1] (with only $\beta
_{1}\neq 0$) and $\widetilde{A}$\ is obtained transposing $A$. The
transformed components $E_{F}^{\prime \mu }$\ are obtained as $E_{F}^{\prime
\mu }=c^{-1}F^{\prime \mu \nu }v_{\nu }$, or explicitly with matrices as
\begin{equation}
\left[
\begin{tabular}{llll}
$0$ & $-F^{\prime 10}$ & $-F^{\prime 20}$ & $-F^{\prime 30}$ \\
$E^{1}$ & $0$ & $-F^{\prime 21}$ & $-F^{\prime 31}$ \\
$\gamma (E^{2}-\beta cB^{3})$ & $\gamma (-\beta E^{2}+cB^{3})$ & $0$ & $%
-F^{\prime 32}$ \\
$\gamma (E^{3}+\beta cB^{2})$ & $\gamma (-\beta E^{3}-cB^{2})$ & $cB^{1}$ & $%
0$%
\end{tabular}%
\right] \cdot \left[
\begin{tabular}{l}
$1$ \\
$0$ \\
$0$ \\
$0$%
\end{tabular}%
\right] =\left[
\begin{tabular}{l}
$0$ \\
$E^{1}$ \\
$\gamma (E^{2}-\beta cB^{3})$ \\
$\gamma (E^{3}+\beta cB^{2})$%
\end{tabular}%
\right] ,  \label{Em4}
\end{equation}%
where the first matrix is the Lorentz transformed $F^{\mu \nu }$, i.e., $%
F^{\prime \mu \nu }$, and the second matrix is $c^{-1}v^{\mu }=\gamma
_{0}^{\mu }$. The components $E_{F}^{\prime \mu }$\ are already written in
equation (\ref{ut}). As seen from (\ref{Em4}) the transformed zeroth
component $E_{F}^{\prime 0}$ is again $=0$, which shows, as previously
stated, that such transformations cannot be the mathematically correct LT;
the LT cannot transform the 4D vector with $E^{0}=0$ into the 4D vector with
$E_{F}^{\prime 0}=0$. Furthermore, it can be simply checked using (\ref{Em4}%
) that for the CBGQs holds
\begin{equation}
E_{F}^{\prime \mu }\gamma _{\mu }^{\prime }\neq E^{\mu }\gamma _{\mu },
\label{em4}
\end{equation}%
where $E_{F}^{\prime \mu }$\ is from (\ref{Em4}).\ This is the same as in (%
\ref{fr}), i.e., it additionally proves that $E_{F}^{\prime \mu }$\ is not
obtained by the mathematically correct LT from $E^{\mu }$.

\emph{Under the mathematically correct LT both} $F^{\mu \nu }$\emph{\ and
the velocity of the observer }$v=c\gamma _{0}$\emph{\ are transformed}. Then
(symbolically)
\begin{equation}
E=c^{-1}F\cdot v\longrightarrow E^{\prime }=c^{-1}F^{\prime }\cdot v^{\prime
}=c^{-1}(AF\widetilde{A})(A^{-1}v)=A(c^{-1}Fv)=AE,  \label{ecm}
\end{equation}%
where, here, $E$, $F$, $v$, $A$, $F^{\prime }$, ... denote matrices. Hence, $%
E^{\prime \mu }$\ can be written as
\begin{equation}
E^{\prime \mu }=c^{-1}F^{\prime \mu \nu }v_{\nu }^{\prime }=c^{-1}(A_{\rho
}^{\mu }F^{\rho \sigma }\widetilde{A}_{\sigma }^{\nu })((A^{-1})_{\nu
}^{\alpha }v_{\alpha })=A_{\rho }^{\mu }(c^{-1}F^{\rho \alpha }v_{\alpha }).
\label{emc}
\end{equation}%
Using the explicit matrices $c^{-1}A^{-1}v$ is given as

\begin{equation}
c^{-1}A^{-1}v=c^{-1}\left[
\begin{tabular}{llll}
$\gamma $ & $\beta \gamma $ & $0$ & $0$ \\
$\beta \gamma $ & $\gamma $ & $0$ & $0$ \\
$0$ & $0$ & $1$ & $0$ \\
$0$ & $0$ & $0$ & $1$%
\end{tabular}%
\right] \cdot \left[
\begin{tabular}{l}
$c$ \\
$0$ \\
$0$ \\
$0$%
\end{tabular}%
\right] =\left[
\begin{tabular}{l}
$\gamma $ \\
$\beta \gamma $ \\
$0$ \\
$0$%
\end{tabular}%
\right]  \label{vc}
\end{equation}%
and $E^{\prime \mu }$\ is $E^{\prime \mu }=c^{-1}F^{\prime \mu \nu }v_{\nu
}^{\prime }$, i.e.,%
\begin{equation}
\left[
\begin{tabular}{llll}
$0$ & $-E^{1}$ & $-F^{2^{\prime }0^{\prime }}$ & $-F^{3^{\prime }0^{\prime
}} $ \\
$E^{1}$ & $0$ & $-F^{2^{\prime }1^{\prime }}$ & $-F^{3^{\prime }1^{\prime }}$
\\
$\gamma (E^{2}-\beta cB^{3})$ & $\gamma (-\beta E^{2}+cB^{3})$ & $0$ & $%
-F^{3^{\prime }2^{\prime }}$ \\
$\gamma (E^{3}+\beta cB^{2})$ & $\gamma (-\beta E^{3}-cB^{2})$ & $cB^{1}$ & $%
0$%
\end{tabular}%
\right] \cdot \left[
\begin{tabular}{l}
$\gamma $ \\
$\beta \gamma $ \\
$0$ \\
$0$%
\end{tabular}%
\right] =\left[
\begin{tabular}{l}
$-\beta \gamma E^{1}$ \\
$\gamma E^{1}$ \\
$E^{2}$ \\
$E^{3}$%
\end{tabular}%
\right] ,  \label{et}
\end{equation}%
where again the first matrix is $F^{\prime \mu \nu }$, as in (\ref{Em4}),\
but the second matrix is the Lorentz transformed 4-velocity of the observer,
i.e., it is given by equation (\ref{vc}). Observe that the same result for $%
E^{\prime \mu }$\ is obtained from $E^{\prime \mu }=A_{\nu }^{\mu }E^{\nu }$,%
\begin{equation}
E^{\prime \mu }=A_{\nu }^{\mu }E^{\nu }=\left[
\begin{tabular}{llll}
$\gamma $ & $-\beta \gamma $ & $0$ & $0$ \\
$-\beta \gamma $ & $\gamma $ & $0$ & $0$ \\
$0$ & $0$ & $1$ & $0$ \\
$0$ & $0$ & $0$ & $1$%
\end{tabular}%
\right] \cdot \left[
\begin{tabular}{l}
$0$ \\
$E^{1}$ \\
$E^{2}$ \\
$E^{3}$%
\end{tabular}%
\right] =\left[
\begin{tabular}{l}
$-\beta \gamma E^{1}$ \\
$\gamma E^{1}$ \\
$E^{2}$ \\
$E^{3}$%
\end{tabular}%
\right] .  \label{an}
\end{equation}%
The components $E^{\prime \mu }$\ are the same as in (\ref{LTE}). This
result clearly shows that the transformations in which \emph{both} $F$ and
the velocity of the observer $v$\ are transformed are the mathematically
correct LT; \emph{under such LT the electric field 4D vector transforms
again only to the electric field 4D vector as any other 4D vector transforms.%
}

As an additional proof of that result it can be simply checked using (\ref%
{an}) that for the CBGQs $E^{\nu }\gamma _{\nu }$, $E^{\prime \nu }\gamma
_{\nu }^{\prime }$, ... again holds the relation (\ref{erc}), $E=E^{\nu
}\gamma _{\nu }=E^{\prime \nu }\gamma _{\nu }^{\prime }=E_{r}^{\nu }r_{\nu
}=E_{r}^{\prime \nu }r_{\nu }^{\prime }$, as for any other CBGQ.\bigskip
\medskip

\noindent \textit{5.2. The electric and magnetic fields as bivectors\bigskip
}

\noindent In [20] the same result about the fundamental difference between
the AT and the correct LT is obtained representing the electric and magnetic
fields by bivectors. The representation by bivectors is used, e.g., in [27,
28] and they derived the AT in which the components of the transformed
electric field bivector are expressed by the combination of components of
the electric and magnetic field bivectors like in (\ref{ee}). In the $\gamma
_{0}$ - frame the electric field bivector $\mathbf{E}_{H}$ is determined
from the electromagnetic field bivector, equation (2) in [20], $\mathbf{E}%
_{H}=(F\cdot \gamma _{0})\gamma _{0}=(1/2)(F-\gamma _{0}F\gamma _{0})$. In
section 5 in [20] the derivation of the AT from [27, 28] is presented. The
space-time split is made and accordingly the space-space components are zero
for the matrix of the electric field bivector $(\mathbf{E}_{H})^{\mu \nu }$,
equation (5) in [20], i.e., $(\mathbf{E}_{H})^{i0}=F^{i0}=E^{i}$, $(\mathbf{E%
}_{H})^{ij}=0$.\ Then, in [27, 28], the same is supposed to hold for the
electric field bivector that is transformed by the AT, equations (18) and
(19)\ in [20]. The transformed electric field bivector $\mathbf{E}%
_{H,at}^{\prime }$\ is not obtained in the way in which all other
multivectors transform, but it is obtained that only F is transformed
whereas $\gamma _{0}$ is not transformed, equation (16) in [20],$\ \mathbf{E}%
_{H,at}^{\prime }=(1/2)[F^{\prime }-\gamma _{0}F^{\prime }\gamma
_{0}]=(F^{\prime }\cdot \gamma _{0})\gamma _{0}$.\ This is the treatment
from [27, 28]. They have not noticed that such transformations cannot be the
correct LT because the LT cannot transform the matrix (5) in [20] in which
the space-space components are zero to the matrix (18) in [20] in which
again the space-space components are zero. \emph{The space-time split is not
a Lorentz covariant procedure.} In section 4 in [20] the derivation of the
correct LT is presented. If the matrix (5) in [20], $(\mathbf{E}_{H})^{\mu
\nu }$, is transformed in the way in which the matrix of any other bivector
transforms under the LT, equation (13) in [20], then the matrix (12) in
[20], $(\mathbf{E}_{H}^{\prime })^{\mu \nu }$,\ is obtained in which the
space-space components are different from zero and \emph{the components} $(%
\mathbf{E}_{H})^{\mu \nu }$ \emph{transform under the LT again to the
components} $(\mathbf{E}_{H}^{\prime })^{\mu \nu }$; \emph{there is no
mixing with the components of the matrix of the magnetic field bivector.} In
general, as shown in [22, 23] the electric and magnetic fields can be
represented by different algebraic objects; vectors, bivectors or their
combination.

\emph{The correct LT always transform the 4D algebraic object representing
the electric field only to the electric field; there is no mixing with the
magnetic field.}\bigskip \bigskip

\noindent \textbf{6. The derivations of the AT of} $\mathbf{E}$ \textbf{and}
$\mathbf{B}$ \textbf{in BT [7]\bigskip }

\noindent As mentioned in the Introduction \emph{the nature of electric and
magnetic fields is discussed}\ in section 1.10 in [7]. There, it is
concluded that \emph{these fields are the 4D fields.} If one applies the LT
to BT's equation (1.109) (it is our equation (\ref{ebv})), e.g., to the
electric field 4D vector then, as discussed above, \emph{both} $F^{\alpha
\beta }$ \emph{and} $w_{\beta }$ (their $w$ is our $v$) have to be
transformed. The equation (\ref{LTE}) would be obtained and equation (\ref%
{erc}) would hold. This is \emph{not} noticed by Blandford and Thorne, [7],
and they believe as all others that their equation (1.113) with the 3D
vectors (the same as equation (11.149) in [1]) is the mathematically correct
\textquotedblleft \textit{Relationship Between Fields Measured by Different
Observers}.\textquotedblright\ Thus, although they deal with 4D GQs they
still consider that in the 4D spacetime, in the same way as in the 3D space,
the 3D vectors are the physical quantities, whereas the 4D quantities are
considered to be only mathematical, auxiliary, quantities. This is visible
in the treatment of the Lorentz force in [7]. In the usual formulations the
physical meaning of 3D vectors $\mathbf{E}$ and $\mathbf{B}$ is determined
by the Lorentz force as a 3D vector $\mathbf{F}_{L}\mathbf{=}q\mathbf{E}+q%
\mathbf{u}\times \mathbf{B}$\ and by Newton's second law $\mathbf{F}=d%
\mathbf{p}/dt$, $\mathbf{p=}m\gamma _{u}\mathbf{u}$. BT start with the
correct equation (1.106) ($dp^{\mu }/d\tau =(q/c)F^{\mu \nu }u_{\nu }$, our
notation), but then instead of to use the decomposition of $F^{\mu \nu }$,
their equation (1.110), our equation (\ref{fm}), they deal with the usual
identification of \emph{the components (in the standard basis) of} $F^{\mu
\nu }$ with \emph{the components of the 3D vectors} $\mathbf{E}$ and $%
\mathbf{B}$, their equation (1.107), our equation (\ref{ieb}), which, as
discussed above, is synchronization dependent and even meaningless in the $%
\{r_{\mu }\}$ basis, see equations (\ref{are}) and (\ref{FEr}). Obviously BT
do not know for the $\{r_{\mu }\}$ basis. Finally they get \textquotedblleft
the familiar Lorentz-force form\textquotedblright\ in terms of \emph{the 3D
vectors} $\mathbf{E}$ and $\mathbf{B}$, their equation (1.108). Thus, the
same as in the usual approaches.

However, in the 4D spacetime, as mentioned above, the Lorentz force $K_{L}$\
is given by equation (\ref{LF}) in terms of $F$ and $u$. Using the
decomposition of $F$ (\ref{E2}) the Lorentz force $K_{L}$\ becomes
\begin{equation}
K_{L}=(q/c)\left[ (1/c)E\wedge v+(IB)\cdot v\right] \cdot u,  \label{lfa}
\end{equation}%
where $u$ is the velocity vector of a charge $q$ (it is defined to be the
tangent to its world line). Note that \emph{there are two velocity vectors in%
} $K_{L}$\ \emph{if it is expressed in terms of fields} $E$ \emph{and} $B$,
because $E$ and $B$ are determined relative to the observer with velocity
vector $v$. If $K_{L}$\ is represented as a CBGQ in the standard basis it is
\begin{equation}
K_{L}=K_{L}^{\mu }\gamma _{\mu }=(q/c)F^{\mu \nu }u_{\nu }\gamma _{\mu
}=(q/c)\{[(1/c)(E^{\mu }v^{\nu }-E^{\nu }v^{\mu })+\varepsilon ^{\lambda \mu
\nu \rho }v_{\lambda }B_{\rho }]u_{\nu }\}\gamma _{\mu },  \label{lf}
\end{equation}%
where $F^{\mu \nu }$\ is from equation (\ref{fm}). In contrast to the usual
expression for the Lorentz force with the 3D fields $\mathbf{E}$ and $%
\mathbf{B}$, $\mathbf{F}_{L}\mathbf{=}q\mathbf{E}+q\mathbf{u}\times \mathbf{B%
}$, the Lorentz force with the 4D fields $E$ and $B$ (\ref{lfa}) or (\ref{lf}%
) contains not only the 4D velocity $u$ of a charge $q$\ but also the 4D
velocity $v$ of the observer who measures 4D fields. It can be simply
checked that for $K_{L}^{\mu }\gamma _{\mu }$\ (\ref{lf}) the relation (\ref%
{klc}) holds
\begin{equation}
K_{L}=K_{L}^{\mu }\gamma _{\mu }=K_{L}^{\prime \mu }\gamma _{\mu }^{\prime
}=K_{Lr}^{\mu }r_{\nu }=K_{Lr}^{\prime \mu }r_{\nu }^{\prime }  \label{klc}
\end{equation}%
as for any other 4D CBGQ. \emph{In the 4D spacetime, the physical meaning of}
$E^{\mu }$ \emph{and} $B^{\mu }$ \emph{is determined by the Lorentz force} $%
K_{L}$ (\ref{lfa}), i.e., $\ K_{L}^{\mu }\gamma _{\mu }$\ (\ref{lf}) \emph{%
and by the 4D expression for Newton's second law}
\begin{equation}
K_{L}^{\mu }\gamma _{\mu }=(dp^{\mu }/d\tau )\gamma _{\mu },\quad p^{\mu
}=mu^{\mu },  \label{pm}
\end{equation}%
$p^{\mu }$ is the proper momentum (components) and $\tau $\ is the proper
time. \emph{All components} $E^{\mu }$ \emph{and} $B^{\mu }$, \emph{thus} $%
E^{0}$ \emph{and} $B^{0}$ \emph{as well, are equally well physical and
measurable quantities by means of the mentioned} $K_{L}^{\mu }$\ (\ref{lf})
\emph{and the 4D expression for Newton's second law} (\ref{pm}) (with $%
K_{L}^{\mu }$\ instead of some arbitrary $K^{\mu }$). Hence, in the 4D
spacetime, contrary to the assertion from [7], the use of the mathematically
correct 4D GQs as in (\ref{lfa}) or (\ref{lf}) cannot lead to
\textquotedblleft the familiar Lorentz-force form.\textquotedblright

Furthermore, BT in [7], state: \textquotedblleft Only after making such an
observer-dependent \textquotedblleft 3+1 split\textquotedblright\ of
spacetime into space plus time do the electric field and magnetic field come
into existence as separate entities.\textquotedblright\ But, as shown above,
in the 4D spacetime \textquotedblleft 3+1 split\textquotedblright\ is
ill-defined. It does not hold in the $\{r_{\mu }\}$ basis and even in the $%
\{\gamma _{\mu }\}$ basis \emph{it is not a Lorentz covariant procedure},
i.e., the 3-surface of simultaneity for one observer (with 4D velocity $w$)
cannot be transformed by the LT into the 3-surface of simultaneity for a
relatively moving inertial observer (with 4D velocity $w^{\prime }$). If for
one observer$\ w^{\mu }=(1,0,0,0)\ $then for a relatively moving inertial
observer it holds that $w^{\prime \mu }=(\gamma ,-\beta \gamma ,0,0)$).
Hence, it cannot be mathematically correct that \emph{both} $E_{w}^{0}=0$
and\ $E_{w^{\prime }}^{0}=0$, but it is necessary $E_{w^{\prime }}^{0}\neq 0$%
, as in (\ref{LTE}) or (\ref{an}). This means that their equation (1.107) is
not correct. It does not follow from equation (1.109), our equation (\ref%
{ebv}) (without unit 4D vectors). Also, equation (1.113) cannot be obtained
by a mathematically correct procedure from equation (1.110). Simply, in the
4D spacetime there is no room for the 3D quantities; an independent physical
reality has to be consistently attributed to the 4D GQs and not to the usual
3D quantities. Obviously, an important statement from Chapter 1 in [7] that
is already mentioned above: \textquotedblleft \emph{We shall state physical
laws, e.g. the Lorentz force law, as geometric, coordinate-free
relationships between these geometric, coordinate free quantities,}%
\textquotedblright\ has to be changed in this way:

\emph{In the 4D spacetime physical laws, e.g. the Lorentz force law, are
geometric, coordinate-free relationships between the 4D geometric,
coordinate free quantities}.

The 3D fields $\mathbf{E}$ and $\mathbf{B}$ and the Lorentz force $\mathbf{F}%
_{L}$ ($\mathbf{F}_{L}=q\mathbf{E}+q\mathbf{u}\times \mathbf{B}$) are also
geometric quantities but in the 3D space, which means that they do not have
well-defined mathematical and physical meaning in the 4D spacetime.

In addition, BT in [7], consider, as almost the whole physics community,
that the Lorentz contraction and the time dilation are the intrinsic
relativistic effects. But, as already mentioned, in [3], [5] and in Appendix
in [2], it is exactly proved that such an opinion is not correct since both
the Lorentz contraction and the time dilation are ill-defined in the 4D
spacetime. Instead of them the 4D GQs, the position 4D vector, the distance
4D vector between two events and the spacetime length have to be used, since
they are properly defined quantities in the 4D spacetime.\bigskip \medskip

\noindent \textit{6.1. Additional comments about the 4D Lorentz
force\bigskip }

Here it is at place to give some additional comments about the Lorentz force
$K_{L}$ (\ref{lfa}) or (\ref{lf}) as a 4D GQ. It is visible from (\ref{lfa})
or (\ref{lf}) that the Lorentz force ascribed by an observer comoving with a
charge, $u=v$, i.e., if the charge and the observer world lines coincide,
then $K_{L}$ is purely electric, $K_{L}=qE$. In the general case when $u$ is
different from $v$, i.e. when the charge and the observer have distinct
world lines, $K_{L}$ (\ref{lfa}) or (\ref{lf}) can be written in terms of $E$
and $B$ as a sum of the $v$ - orthogonal part, $K_{L\perp }$ ($K_{L\perp
}\wedge v=0$) and $v$ - parallel part, $K_{L\parallel }$ ($K_{L\parallel
}\cdot v=0$). As the CBGQs they are%
\begin{eqnarray}
K_{L} &=&K_{L\perp }+K_{L\parallel },\quad K_{L\perp }=(q/c^{2})[(v^{\nu
}u_{\nu })E^{\mu }+\varepsilon ^{\lambda \mu \nu \rho }v_{\lambda }u_{\nu
}cB_{\rho }]\gamma _{\mu },  \notag \\
K_{L\parallel } &=&(q/c^{2})[-(E^{\nu }u_{\nu })v^{\mu }]\gamma _{\mu }.
\label{kop}
\end{eqnarray}%
Speaking in terms of the prerelativistic notions one can say that in the
approach with the vectors $E$ and $B$ the $v$ - orthogonal part, $K_{L\perp
} $, from (\ref{kop}) plays the role of the usual Lorentz force lying on the
3D hypersurface orthogonal to $v$, whereas $K_{L\parallel }$ from (\ref{kop}%
) is related to the work done by the field on the charge. This can be seen
specifying (\ref{kop}) to the $\gamma _{0}$ - frame, $v=c\gamma _{0}$, in
which $E^{0}=B^{0}=0$. In the $\gamma _{0}$ - frame it is possible to
compare the 4D vector $K_{L}$\ with the usual 3D Lorentz force, $\mathbf{F}%
_{L}\mathbf{=}q\mathbf{E}+q\mathbf{u}\times \mathbf{B}$, which yields%
\begin{eqnarray}
K_{L}^{0}\gamma _{0} &=&K_{L\parallel }^{0}\gamma
_{0}=-(q/c)E^{i}u_{i}\gamma _{0},\quad K_{L\perp }^{0}=0,  \notag \\
K_{L}^{i}\gamma _{i} &=&K_{L\perp }^{i}\gamma _{i}=q((E^{i}+\varepsilon
^{0ijk}u_{j}B_{k})\gamma _{i},\quad K_{L\parallel }^{i}\gamma _{i}=0
\label{koi}
\end{eqnarray}%
It is visible from (\ref{koi}) that $K_{L}^{0}$\ is completely determined by
$K_{L\parallel }$,\ whereas the spatial components $K_{L}^{i}$\ are
determined by $K_{L\perp }$.\ However, as already mentioned several times,
in this 4D geometric approach \emph{only both parts taken together, i.e.,
the whole }$K_{L}=K_{L\perp }+K_{L\parallel }$\emph{\ does have a definite
physical meaning and it defines the 4D Lorentz force both in the theory and
in experiments. }

In section 2.5 in [14], under the title \textquotedblleft The Lorentz force
and the motion of charged particle in the electromagnetic field $F$%
\textquotedblright\ the definition of $K_{L}$ in terms of $F$ is exclusively
used ($K_{L}=(q/c)F\cdot u$) without introducing the electric and magnetic
fields. Observe that the 4D GQs $K$ ($K_{L}$), $p$, $u$ transform in the
same way, like any other 4D vector, i.e., according to the LT and not
according to the awkward AT of the 3D force $\mathbf{F}$, e.g., equations
(12.66) and (12.67) in [4], and the 3D momentum $\mathbf{p}$, i.e., the 3D
velocity $\mathbf{u}$. In [29], under the title \textquotedblleft Four
Dimensional Geometric Quantities versus the Usual Three-Dimensional
Quantities: The Resolution of Jackson's Paradox,\textquotedblright\ it is
shown that only with the use of the 4D Lorentz force (\ref{lfa}), (\ref{lf})
or (\ref{kop}), the torque bivector $N=(1/2)N^{\mu \nu }\gamma _{\mu }\wedge
\gamma _{\nu }$, $N^{\mu \nu }=x^{\mu }K_{L}^{\nu }-x^{\nu }K_{L}^{\mu }$
and the angular momentum bivector $M=(1/2)M^{\mu \nu }\gamma _{\mu }\wedge
\gamma _{\nu }$, $M^{\mu \nu }=m(x^{\mu }u^{\nu }-x^{\nu }u^{\mu })$ there
is no apparent electrodynamic paradox with the torque and that the principle
of relativity is naturally satisfied. The mentioned paradox is described in
[30]\ and it consists in the fact that there is a 3D torque $\mathbf{N}$ and
thus $d\mathbf{L}/dt$ ($\mathbf{N}=d\mathbf{L}/dt$) in one inertial frame,
but no 3D angular momentum $\mathbf{L}^{\prime }$ and no 3D torque $\mathbf{N%
}^{\prime }$ in another relatively moving inertial frame. Similar
electrodynamic paradoxes with the 3D torque appear in the Trouton-Noble
paradox, see, e.g., [31], and the \textquotedblleft charge-magnet
paradox\textquotedblright\ [32]. Using the above mentioned 4D GQs, 4D
Lorentz force, the torque and angular momentum bivectors it is explicitly
shown in [33], [14], for the Trouton-Noble paradox and [34], [2] for
Mansuripur's paradox that there is no paradox and consequently there is no
need for some \textquotedblleft resolutions\textquotedblright\ of the
paradoxes, e.g., by the introduction of the Einstein-Laub force, [32], or by
the introduction of some \textquotedblleft hidden\textquotedblright\
quantities, e.g., [35].\ \bigskip \bigskip

\noindent \textbf{7. The shortcomings\ in the derivations of the AT of} $%
\mathbf{E}$ \textbf{and} $\mathbf{B}$

\textbf{and in the treatment of the angular momentums in KS [8]}\textit{%
\bigskip }

\noindent \textit{7.1. The shortcomings\ in the derivations of the AT of} $%
\mathbf{E}$\textit{\ and} $\mathbf{B\bigskip }$

\noindent Similar mistakes as in BT [7] are made by Klajn and Smoli\'{c}
(KS) in section 3 in [8]. KS [8] use the tensor formalism with the abstract
index notation but, nevertheless, they consider as in [7] that the 3D
vectors are well-defined physical quantities in the 4D spacetime whereas the
4D quantities are only mathematical, auxiliary, quantities. In the first
part of section 3 in [8] they derive the transformations of the 3D $\mathbf{E%
}$ and $\mathbf{B}$, their equations (25) and (26), in the same way as in
[1]. The shortcomings of such a derivation are discussed in detail in our
section 1, the objections 1), 2) and in section 2.1, the objections 3), 4)
and 5). As in [7], KS [8] also know \emph{only} for the standard basis and
not for the $\{r_{\mu }\}$ basis in which, according to equations (\ref{are}%
) and (\ref{FEr}), the usual identification, equation (24) in [8], i.e., our
equation (\ref{ieb}), is meaningless even in their specific inertial
reference frame $\mathcal{R}$, what is the $\gamma _{0}$ - frame in our
notation. For the same reasons, contrary to their assertion, it is not true
that the identifications (\ref{eb2}) hold for a relatively moving inertial
frame $\mathcal{R}^{\prime }$ too. As already discussed at the end of
section 1, their $F_{ab}$, our $F$, is represented as in (\ref{fc}) and it
contains not only components but a basis as well, which means that their
relation $F_{ab}\rightarrow F_{\mu \nu }$\ is not mathematically correct. In
the second part of section 3 in [8] they deal, as they say, with
\textquotedblleft an alternative approach\textquotedblright\ in which the
observers which measure the electric and magnetic fields are explicitly
introduced.

The mathematical incorrectness of their derivation can be best seen, e.g.,
from their discussion at the end of section 3 and equations (34) - (37) in
[8]. They, KS, construct the electric 4-vectors, in the same way as it is
made by BT in [7]. In [8] it is assumed that if the 4-velocity of the
observer in $\mathcal{R}$\ is in the $\gamma _{0}$ direction, $v=c\gamma
_{0} $,\ and consequently $E=F\cdot \gamma _{0}$ with the components $E^{\mu
}=(E^{0}=0$, $E^{1}$, $E^{2}$, $E^{3})$, then the same relations must hold
for a relatively moving inertial observer, $v^{\prime }=c\gamma _{0}^{\prime
}$ and $E^{\prime \mu }=(E^{\prime 0}=0$, $E^{\prime 1}$, $E^{\prime 2}$, $%
E^{\prime 3})$. In their notation, for the observer $o$ with $o^{\mu }=(c,%
\mathbf{0})$, $E^{a}(o)=F^{ab}o_{b}$ so that $E^{\mu }(o)=(0,\mathbf{E)}$
and it is supposed that the same holds for the observer $o^{\prime }$, $%
E^{a}(o^{\prime })=F^{ab}o_{b}^{\prime }$ so that $E^{\mu ^{\prime
}}(o^{\prime })=(0,\mathbf{E}^{\prime }\mathbf{)}$. In [8] it is stated:
\textquotedblleft The 4-vector $E^{a}(o)$ is related to the electric field 3
- vector as measured by $o$, and the same holds for $E^{a}(o^{\prime })$ and
the observer\ $o^{\prime }$.\textquotedblright\ We remark that the same
relation has to hold for the observers $S^{\prime \prime }$, $S^{\prime
\prime \prime }$ ($o^{\prime \prime }$, $o^{\prime \prime \prime }$) etc.,
since \emph{it is the definition of the vector} $E$. However, it is not
understood by KS that $E^{\prime }$\ and $v^{\prime }$ from $E^{\prime
}=F\cdot v^{\prime }/c=F\cdot \gamma _{0}^{\prime }$ \emph{are not }the
Lorentz transforms and they have nothing to do with the LT of $E$\ and $v$
from $E=F\cdot v/c$. The reason is that $\gamma _{0}$\ is transformed by the
LT as in equation (\ref{g0}),%
\begin{equation}
\gamma _{0}=\gamma (\gamma _{0}^{\prime }-\beta \gamma _{1}^{\prime }).
\label{g0}
\end{equation}%
As it is discussed in section 5 the unit vector in the time direction $%
\gamma _{0}$\ (from $v=c\gamma _{0}$,$\ E=F\cdot \gamma _{0}$) for the
observer $S$ \emph{is not transformed }by the LT into the unit vector in the
time direction $\gamma _{0}^{\prime }$\ for the observer $S^{\prime }$ (from
$v^{\prime }=c\gamma _{0}^{\prime }$, $E^{\prime }=F\cdot \gamma
_{0}^{\prime }$), which means that if $v$ is in the $\gamma _{0}$\ direction
then, as said above, $v^{\prime }$ \emph{cannot be} in the $\gamma
_{0}^{\prime }$ direction. One can take any observer as the starting one for
which $E$\ is defined as in $E=F\cdot v/c$ and then to find the electric
field vector $E^{\prime }$\ for a relatively moving observer $S^{\prime }$\
one has to perform the active LT of that $E$\ in a mathematically correct
way, i.e., for the active LT as in equations (\ref{Lt}) - (\ref{el}). In
addition, it is worth mentioning that their notation $E^{\mu }(o)=(0,\mathbf{%
E)}$, $E^{\mu ^{\prime }}(o^{\prime })=(0,\mathbf{E}^{\prime }\mathbf{)}$,
etc. is not correct not only because the temporal component in $E^{\mu
^{\prime }}(o^{\prime })$\ cannot be zero, but for other reasons too.
Firstly, usually $\mathbf{E}$\ denotes the 3D electric field in which the
components $E_{x,y,z}$ are multiplied by the unit 3D vectors $\mathbf{i}$, $%
\mathbf{j}$, $\mathbf{k}$, whereas in $E^{\mu }(o)$\ they have to be $E^{1}$%
, $E^{2}$, $E^{3}$, which in the geometric quantity $E^{a}(o)$\ would need
to be multiplied by the spatial unit 4D vectors. In the 4D spacetime there
are no 3D vectors. Moreover, as already said, the standard basis is
implicitly assumed in the whole paper [8]. But, an observer can use
different bases. Particularly, if the $\{r_{\mu }\}$ basis is used then, as
seen from (\ref{ei}),\ the temporal component of $E^{\mu }(o)$\ in the $%
\{r_{\mu }\}$ basis, $E_{r}^{0}\neq 0$, even if it is $=0$ in the standard
basis. This is not taken into account in their formulation and with their
notation.

Let us explain the shortcomings and misconceptions in their derivations in
another way too. The whole their reasoning is clearly visible from their
equation (35). In that equation, in their notation, they have on the r.h.s. $%
E^{\mu }=(E^{0}=0$, $E^{1}$, $E^{2}$, $E^{3})$ and also on the l.h.s. $%
E^{\prime \mu ^{\prime }}=(E^{\prime 0^{\prime }}=0$, $E^{\prime 1^{\prime
}} $, $E^{\prime 2^{\prime }}$, $E^{\prime 3\prime })$, i.e., the temporal
component of the electric 4D vector is taken to be zero for both relatively
moving inertial observers $o$ and $o^{\prime }$. Then it is stated in [8]
that the only LT that satisfies the equation (35) is the 3-rotation
transformation. However, \emph{they erroneously consider that in both
relatively moving inertial frames the temporal components have to be zero.}
This is completely equivalent to the treatment from [7] in which it is
supposed that the \textquotedblleft 3+1 split\textquotedblright\ of the
spacetime into space + time holds in both relatively moving inertial frames,
i.e., that it is a Lorentz covariant procedure. As already explained several
times, in the 4D spacetime the physical quantities are the 4D geometric
quantities and not the 3D vectors, which means that the LT will necessary
transform the electric 4D-vector with $E^{0}=0$ into the electric 4D-vector
with $E^{\prime 0\prime }\neq 0$. In the 4D spacetime, as stated above, all
components of the 4D vectors $E$ and $B$ including $E^{0}$ and $B^{0}$ are
equally well physical and measurable quantities by means of the equations (%
\ref{lf}) and (\ref{pm}). Their, [8], equation (35) has to have on the
l.h.s. $E^{\prime 0^{\prime }}\neq 0$. Only in that case it will be a
mathematically correct LT (boost) of the 4D electric vector from the $\gamma
_{0}$ - frame (the r.h.s. of (35)) and the components will be given by
equation (\ref{LTE}). Thus, the mathematically incorrect equations (34) and
(35) in [8] has to be replaced with our mathematically correct LT (boost) (%
\ref{et}) and (\ref{an}), i.e., (\ref{tec}). In that case, as stated at the
end of section 5.1, the relation (\ref{erc}) holds as for any other 4D
vector. From the mathematical viewpoint under the passive LT both the
components and the basis are Lorentz transformed but the 4D vector $E$\
remained unchanged. The 4D rotation of the basis is performed, e.g., for the
standard basis, $\gamma _{\mu }\rightarrow \gamma _{\mu }^{\prime }$. The
components of that unchanged $E$\ are determined relative to that new basis,
$E^{\mu }\rightarrow E^{\prime \mu }$. Hence, in (\ref{tec}) as in (\ref{an}%
) $E^{\prime \mu }=A_{\nu }^{\mu }E^{\nu }$. In contrast to the statement
from [8], the components $E^{\mu }$ and $E^{\prime \mu }$\ refer to the
measurements by the observers in two relatively moving inertial frames of
reference, two different bases $\gamma _{\mu }$ and $\gamma _{\mu }^{\prime
} $. The vector $E=E^{\mu }\gamma _{\mu }$\ \emph{is} a genuine 4D vector.
KS [8] do not properly differ between the passive LT and the active LT. As
explained above, section 4.1, under the active LT $E$ is transformed into a
new electric field vector $E^{\prime }$, (\ref{aec}); the components are
changed, (\ref{LTE}), but the basis remains unchanged as in (\ref{eb}). From
the physical viewpoint the measurements are made by the observers in one
frame, one basis, but they are made on two different 4D vectors $E$ and $%
E^{\prime }$, which are connected by the active LT as in (\ref{eb}). It is
visible from (\ref{eb}) and (\ref{an}) that the components of the new vector
$E^{\prime }$\ in the old basis are the same as the components of the old
vector $E$\ in the new basis $\gamma _{\mu }^{\prime }$, as it has to be.
This is not unerstood by KS [8]. In [8] it is also used an unusual and in
some way an awkward notation with primed quantities and primed indices.
Instead of such a strange notation they could simply use the primed
components and bases.

The title of [8] is \textquotedblleft Subtleties of invariance, covariance
and observer independence.\textquotedblright\ From the above discussion it
can be concluded that, contrary to the assertions from section 3 in [8], the
correct LT of the electric field are our equations, (\ref{Lt}) - (\ref{el}),
i.e., with matrices, (\ref{et}) and (\ref{an}). The \textquotedblleft
subtle\textquotedblright\ point that is not understood by KS [8] is that the
LT are properly defined on the 4D spacetime and they cannot transform three
spatial components for one observer again into three spatial components for
relatively moving observer. In other words, \emph{in the 4D spacetime the 4D
vectors are properly defined and not the 3D vectors.} Hence, in the 4D
spacetime the 3-rotation transformation is meaningless and it has nothing to
do with the mathematically correct LT (the rotation in the 4D spacetime)
since the 3D vectors are not well-defined quantities.

The fact that in [8] the 3D $\mathbf{E}$ and $\mathbf{B}$ are considered as
well-defined physical quantities in the 4D spacetime causes an incorrect
expression for the Lorentz force law, their equation (33), $d(mu^{a})/d\tau
=qF_{b}^{a}u^{b}\equiv qE^{a}(u)$.\ The correct expression for the Lorentz
force is $qF_{b}^{a}u^{b}$, but it is completely incorrect to argue that it
is $\equiv qE^{a}(u)$, where : \textquotedblleft $E_{a}(u)$ stands for the
combination of both electric and magnetic 3D vectors (the familiar 3D vector
representation of Lorentz's law).\textquotedblright\ If $E_{a}(u)$\ is
expressed with 3D vectors how then it can be identical to the 4D vector $%
qF_{b}^{a}u^{b}$.\ In the 4D spacetime the mathematically correct
formulation of the Lorentz force law is given in our section 6 by equations (%
\ref{lfa}) or (\ref{lf}), i.e., (\ref{kop}) and (\ref{koi}).

Observe also an important difference between our equations (\ref{E2}), (\ref%
{E1}) and equations (27), (30) in [8]. In our formulation the starting
equation for the introduction of the 4D $E$ and $B$ is equation (\ref{E2}),
i.e., a mathematical theorem that holds for\emph{\ any antisymmetric tensor
of the second rank}. In that theorem the 4D vector $v$\ is a \emph{time-like}
4D vector, which means that it is not necessary in the time direction. $E$
and $B$ are the space-like vectors given by equation (\ref{E1}). It is not
so in [8]. They first define $E^{a}(o)$ and $B^{a}(o)$ by equation (27) in
which $o_{b}$\ \emph{is explicitly in the time-direction} ($o^{\mu }=(c,%
\mathbf{0})$). Then, they construct the Faraday tensor $F_{ab}$\ using such $%
o_{b}$.\ It corresponds to the case that it is chosen $v=c\gamma _{0}$ in
our (\ref{E2}) and (\ref{E1}). However, (\ref{E2}) and (\ref{E1}) hold in
the same measure if $v$\ is not $=c\gamma _{0}$, but it is obtained by the
LT from $c\gamma _{0}$, i.e., $v=c(\gamma \gamma _{0}-\beta \gamma \gamma
_{1})$ That $v$\ is not in the time direction, but it is still a time-like
4D vector, $v^{2}$ is again $=c^{2}$.\ Their, [8], definitions (27) and (30)
with $o^{\mu }=(c,\mathbf{0})$\ are the real cause of all other
mathematicall incorrectnesses in [8], which are discussed in this section.
Instead of (27) and (30) from [8] one has to use (\ref{E2}) and (\ref{E1})
and it has to be in that order. In addition, their, [8], reference [12] is
not correct. The title of that paper, reference [21] here, is: The proof
that the standard transformations of E and B are not the Lorentz
transformations.\bigskip

\noindent \textit{7.2. The shortcomings in the treatment of the angular
momentums in }[8]\textit{\bigskip }

\noindent There are even more mathematical incorrectnesses in the treatment
of the angular momentums in section 4 in [8]. They start the consideration
with equation (38) in which the components (implicitly taken in the standard
basis) $J_{\mu \nu }$\ of the angular momentum tensor $J_{ab}$\ in $\mathcal{%
R}$\ (our $\gamma _{0}$ - frame) are\ identified with the components of two
3D vectors $\mathbf{K}$\ and $\mathbf{J}$. As they say: \textquotedblleft $%
\mathbf{K}$\ is the boost 3D vector describing the movement of the
particle's center of mass, while $\mathbf{J}$ is the angular momentum 3D
vector.\textquotedblright\ In the usual covariant approaches, e.g., [1],
[4], [36] the 3D vectors are considered as primary physical quantities that
determine the components $F_{\mu \nu }$\ of the electromagnetic field
tensor. In the same way in [8] the components of the 3D vectors $\mathbf{K}$%
\ and $\mathbf{J}$ are considered as primary physical quantities that
determine the components $J_{\mu \nu }$\ of the angular momentum tensor.
Firstly, as already discussed several times, such an identification of \emph{%
the components (in the standard basis) of} $J_{\mu \nu }$\ with \emph{the
components of the 3D vectors} $\mathbf{K}$\ and $\mathbf{J}$, their equation
(38), is synchronization dependent and even meaningless in the $\{r_{\mu }\}$
basis. The objections 1), 2) from our section 1 and 3), 4) from section 2.1
hold in the same measure for their treatment of the angular momentums.
However, in this case, there are some additional objections. The first one
refers to the physical interpretation of the components of $J_{\mu \nu }$\
in their equation (38). According to their interpretation the three
components of $\mathbf{K}$, i.e., the \textquotedblleft
time-space\textquotedblright\ components of $J_{\mu \nu }$, are not the
angular momentum components as are the \textquotedblleft
space-space\textquotedblright\ components $J_{x,y,z}$. This is also visible
from their equation (39), which, as they state, defines the angular momentum
4-vector. If that equation would be writen with CBGQs in their specific
reference frame $\mathcal{R}$\ then the components of that angular momentum
4-vector would be $(0,J_{x},J_{y},J_{z})$\ ($J^{0}=0,\
J^{i}=(1/2)\varepsilon ^{0ijk}J_{jk}$). Only the \textquotedblleft
space-space\textquotedblright\ components of $J_{\mu \nu }$ define the
spatial components of the angular momentum 4-vector $J^{a}(o)$.\ The
temporal component of that 4D-vector is $=0$. However, if one uses the $%
\{r_{\mu }\}$ basis instead of the standard basis then, in the same way as
in relations (\ref{are}) and (\ref{FEr}), one would get that, e.g., the
\textquotedblleft time-space\textquotedblright\ component $K_{x,r}$\ of the
component form of the angular momentum four-tensor in the $\{r_{\mu }\}$
basis, $J_{\mu \nu (r)}$, would be expressed as the combination of the
\textquotedblleft time-space\textquotedblright\ component $K_{x}$\ and the
\textquotedblleft space-space\textquotedblright\ components $J_{y}$ and $%
J_{z}$\ of the same angular momentum four-tensor in the $\{\gamma _{\mu }\}$
basis, i.e., that one whose components are given by equation (38) in [8],\
\begin{equation}
K_{xr}=K_{x}+J_{z}-J_{y}.  \label{kj}
\end{equation}%
This is the reason why we use the quotation marks in \textquotedblleft
time-space\textquotedblright\ and \textquotedblleft
space-space.\textquotedblright\ Furthermore, if the LT of the components $%
J_{\mu \nu }$\ from equation (38) in [8] is performed and the same
identification is used in the relatively moving inertial frame of reference $%
\mathcal{R}^{\prime }$\ then the AT of the components of the 3D vectors $%
\mathbf{K}$\ and $\mathbf{J}$ are obtained

\begin{eqnarray}
J_{x}^{\prime } &=&J_{x},\ J_{y}^{\prime }=\gamma (J_{y}+\beta K_{z}),\
J_{z}^{\prime }=\gamma (J_{z}-\beta K_{y}),  \notag \\
K_{x}^{\prime } &=&K_{x},\ K_{y}^{\prime }=\gamma (K_{y}-\beta J_{z}),\
K_{z}^{\prime }=\gamma (K_{z}+\beta J_{y}).  \label{te}
\end{eqnarray}%
As can be seen from (\ref{te}) these transformations are the same as the AT
for $B_{i}$ and $E_{i}$, respectively. Here, they are written for the motion
along the $x^{1}$ axis. \emph{The essential point is that in }(\ref{te})
\emph{the transformed components} $J_{i}^{\prime }$\emph{\ are expressed by
the mixture of components,} $J_{k}$\emph{,} $K_{k}$ and vice versa. The
above relation for $K_{xr}$\ (\ref{kj}) and the relations (\ref{te}) clearly
show that it is not correct to consider that only three \textquotedblleft
space-space\textquotedblright\ components of $J_{\mu \nu }$ implicitly taken
in the standard basis are the components of the physical angular momentum.
\emph{From the mathematical viewpoint all six independent components of }$%
J_{\mu \nu }$\emph{\ are completely equivalent and they necessarily have to
have the same physical interpretation.} Strictly speaking the components
taken alone are not physical. In this case the physical quantity is the
angular momentum four tensor $J_{ab}$\ as an abstract 4D GQ or its
representation in some basis the 4D CBGQ that contains not only the
components as in equation (38) in [8] but the chosen basis as well.\ In
equations (40) and (41) they, KS in [8], define the orbital angular momentum
$L^{a}$\ and the spin $S^{a}$, respectively. Then they get equation (42) in
which the total angular momentum is written as the sum of the orbital
angular momentum and the spin 4-vector. Observe that again only the
\textquotedblleft space-space\textquotedblright\ part of $J_{ab}$\ is used
to define $L^{a}$\ and $S^{a}$. In order to get that $J^{a}$\ can be written
as a sum of $L^{a}$\ and $S^{a}$,\ their equation (42),\ they define $L^{a}$%
\ in such a way that it contains both, the 4-velocity of the observer $o_{b}$%
\ and the 4-velocity of the\ particle $u_{b}$, their equation (40). Hence it
is not correct to write $L^{a}(o)$\ since it depends on the particle's
4-velocity $u$\ as well.\ Even in the $\mathcal{R}$\ frame in which $o^{\mu
}=(c,0,0,0)$\ the temporal component of $L^{\mu }$\ will be different from
zero and one cannot get the usual expression for the spatial components of
the orbital angular momentum.\ Hence, e.g., $J_{x}$\ from their equation
(38) is not equal to the sum of the usual $L_{x}$\ and $S_{x}$.\ Similarly,
in the treatment of the spin, equations (43) - (45), KS [8] consider that
the spin 3D vector $\mathbf{s}$\ is a well-defined physical quantity in the
4D spacetime and\ that it transforms according to the transformations given
by their equation (45) (equation (11.159) in [1]), which are typical AT of
the 3D vectors.

The treatment of the angular momentums in [8] is very similar to the
treatment of the angular momentum and torque in Jackson's paper [30]. There,
[30], Jackson deals with the usual covariant definition of the angular
momentum four-tensor (orbital) $M^{\mu \nu }=x^{\mu }p^{\nu }-x^{\nu }p^{\mu
}$. The \emph{components} $L_{i}$ of the \emph{3D }orbital angular momentum $%
\mathbf{L}=\mathbf{r}\times \mathbf{p}$\ are identified with the
\textquotedblleft space-space\textquotedblright\ \emph{components} of $%
M^{\mu \nu }$ and the \emph{components} $K_{i}$ of another \emph{3D vector }$%
\mathbf{K}$ are identified with the three \textquotedblleft
time-space\textquotedblright\ \emph{components} of $M^{\mu \nu }$. In [30],
in contrast to [8], it is not given any physical interpretation for $K_{i}$.
It is assumed that $L_{i}$ and $K_{i}$ transform as the \textquotedblleft
space-space\textquotedblright\ and \textquotedblleft
time-space\textquotedblright\ components respectively of the usual covariant
angular momentum four-tensor $M^{\mu \nu }$, see [30] and section 3 in the
first paper in [29]. These AT of the components of $\mathbf{L}$ are the same
as the AT of\ $\mathbf{J}$ in (\ref{te}) but with $L_{i}$ replacing $J_{i}$,
which means that \emph{the transformed components} $L_{i}^{\prime }$\emph{\
are expressed by the mixture of components,} $L_{k}$\emph{,} $K_{k}$ and
vice versa. The same situation happens with the 3D torque $\mathbf{N}$ and
the torque bivector $N=(1/2)N^{\mu \nu }\gamma _{\mu }\wedge \gamma _{\nu }$
in the above mentioned considerations of different electrodynamic paradoxes,
[31, 32]. In [30], and also in, e.g., [31, 32, 35], only the
\textquotedblleft space-space\textquotedblright\ components of $M^{\alpha
\beta }$ ($L_{i}$) and $N^{\alpha \beta }$\ ($N_{i}$) are considered to be
the physical angular momentum and torque respectively, because they are
associated with actual rotation in the 3D space of the object. On the other
hand, the \textquotedblleft time-space\textquotedblright\ components of $%
M^{\alpha \beta }$ ($K_{i}$) and $N^{\alpha \beta }$\ (let us denote them as
$R_{i}$) are \emph{not} considered to be of the same physical nature as $%
L_{i}$ and $N_{i}$. In all usual treatments it is considered that $K_{i}$
and $R_{i}$ are not the physical angular momentum and torque respectively,
because they are \emph{not} associated with any overt rotation in the 3D
space of the object, see, particularly, the paper by Griffiths and Hnizdo in
[35] and Jackson's paper [30]. However, as already discussed above, the
relations (\ref{kj}) and (\ref{te}) reveal that such usual interpretation of
the components of $M^{\alpha \beta }$ and $N^{\alpha \beta }$\ is apparently
incorrect; how it can be physically acceptable that in the relation, e.g., $%
L_{y}^{\prime }=\gamma (L_{y}+\beta K_{z})$, $L_{y}$ and $L_{y}^{\prime }$
are the components of a physical angular momentum, whereas it is not so with
$K_{z}$. The same objection refers to the treatment of the angular momentums
in [8]. \bigskip \bigskip

\noindent \textbf{8. Briefly about the mathematically correct 4D angular
momentums}\textit{\bigskip }

\noindent In contrast to treatment of the angular momentums in [8], the
mathematically correct definitions with the 4D GQs of the orbital angular
momentum bivector are given, e.g., in section 2 in the first paper in [29]
(section 4 in the second paper)%
\begin{equation}
M=x\wedge p,\quad M=(1/2)M^{\mu \nu }\gamma _{\mu }\wedge \gamma _{\nu },\
M^{\mu \nu }=x^{\mu }p^{\nu }-x^{\nu }p^{\mu },  \label{m2}
\end{equation}%
in connection with the discussion of Jackson's paradox and also in [33] and
[34] in the mathematically correct treatment of the Trouton-Noble paradox
and Mansuripur's paradox respectively. The same definitions but in the
tensor formalism with the abstract index notation are given in section 2 in
[10]. In a complete analogy with the decomposition of $F$ into $E$, $B$ and $%
v$, equations (\ref{E2}) - (\ref{ebv}),\ the mathematical theorem from
section 3.3 can be used for the decomposition of the bivector $M$\ into two
4D vectors $M_{s}$ and $M_{t}$ and $v$, the 4D velocity vector of a family
of observers who measures $M$

\begin{eqnarray}
M &=&(v/c)\wedge M_{t}+(v/c)\cdot (M_{s}I),  \notag \\
M_{t} &=&(v/c)\cdot M,\quad M_{s}=I(M\wedge v/c),  \label{lt}
\end{eqnarray}%
whith the condition
\begin{equation}
M_{s}\cdot v=M_{t}\cdot v=0.  \label{cn}
\end{equation}%
Only three components of $M_{s}$ and three components of $M_{t}$ are
independent since $M$ is antisymmetric. If\emph{\ }$M$,$\ M_{s}$ and $M_{t}$
are written as CBGQs in the $\{\gamma _{\mu }\}$ basis then there components
are
\begin{eqnarray}
M^{\mu \nu } &=&(1/c)[(v^{\mu }M_{t}^{\nu }-v^{\nu }M_{t}^{\mu
})+\varepsilon ^{\mu \nu \rho \sigma }M_{s\rho }v_{\sigma }]  \notag \\
M_{s}^{\nu } &=&(1/2c)\varepsilon ^{\alpha \beta \mu \nu }M_{\alpha \beta
}v_{\mu },\ M_{t}^{\nu }=(1/c)M^{\mu \nu }v_{\mu }.  \label{lg}
\end{eqnarray}%
Similarly as for $E$ and $B\ $it can be concluded from (\ref{lt}) - (\ref{lg}%
) that both $M_{s}$ and $M_{t}$ depend not only on $M$ but also on $v$.
Hence, it can be said that the bivector $M$\ is the primary quantity for the
angular momentums. \emph{Both vectors} $M_{s}$ \emph{and} $M_{t}$ \emph{are
physical angular momentums} which contain the same physical information as
the bivector $M$ \emph{only when they are taken together. }In the $\gamma
_{0}$ - frame $v^{\mu }=(c,0,0,0)$, $M_{s}^{0}=M_{t}^{0}=0$ and \emph{only
the spatial components} $M_{s}^{i}$ and $M_{t}^{i}$ remain, $%
M_{t}^{i}=M^{i0} $, $M_{s}^{i}=(1/2)\varepsilon ^{0ijk}M_{jk}$; $%
M_{s}^{1}=M^{23}=x^{2}p^{3}-x^{3}p^{2}$, $M_{s}^{2}=M^{31}$, $%
M_{s}^{3}=M^{12}$. Therefore $M_{s}$ can be called the \textquotedblleft
space-space\textquotedblright\ angular momentum and $M_{t}$ the
\textquotedblleft time-space\textquotedblright\ angular momentum. $M_{s}^{i}$
and $M_{t}^{i}$ correspond to the components of $\mathbf{L}$ and $\mathbf{K}$
that are introduced, e.g., in [30] and discussed in the preceding section.
However, as already mentioned, Jackson [30], as all others, considers that
only the 3D $\mathbf{L}$ is a physical quantity whose components transform
according to equation (11) in [30], i.e., equation (\ref{te}) here but with $%
L_{i}$ replacing $J_{i}$. In contrast to it the 4D vectors $M_{s}$ and $%
M_{t} $ transform under the LT as any other 4D vectors transform, i.e., the
components in the standard basis transform like in equation (\ref{el}).
Under the active LT, e.g., the 4D vector $M_{s}$ transforms again into the
\textquotedblleft space-space\textquotedblright\ angular momentum $%
M_{s}^{\prime }$ and there is no mixing with $M_{t}$.

It is shown in [37, 38, 10, 2] that the same consideration as for the
orbital angular momentum can be applied to the \emph{intrinsic} angular
momentum. The primary quantity \emph{with} \emph{definite physical reality}
for the \emph{intrinsic} angular momenta is the spin bivector $\mathcal{S}$
(four-tensor $S^{ab}$ in [37, 38, 10]), which, as in (\ref{lt}) - (\ref{lg}%
),\ can be decomposed into the usual \textquotedblleft
space-space\textquotedblright\ intrinsic 4D angular momentum vector $S$, the
\textquotedblleft time-space\textquotedblright\ 4D intrinsic angular
momentum vector $Z$ and the unit time-like 4D vector $u/c$, where $u$ is the
4D velocity vector of the particle%
\begin{eqnarray}
\mathcal{S} &=&(1/c)[Z\wedge u+(SI)\cdot u],  \notag \\
Z &=&\mathcal{S}\cdot u/c,\quad S=I(\mathcal{S}\wedge u),  \label{e}
\end{eqnarray}%
equation (58) in [2], or with $S^{ab}$, equation (8) in [10]. It holds that $%
Z\cdot u=S\cdot u=0$; only three components of $Z$ and three components of $%
S $ are independent since $\mathcal{S}$ is antisymmetric. $S$ and $Z$ depend
not only on $\mathcal{S}$ but on $u$ as well. Only in the particle's rest
frame, the $K^{\prime }$ frame, and the $\{\gamma _{\mu }^{\prime }\}$
basis, $u=c\gamma _{0}^{\prime }$ and $S^{\prime 0}=Z^{\prime 0}=0$, $%
S^{\prime i}=(1/2c)\varepsilon ^{0ijk}\mathcal{S}_{jk}^{\prime }$, $%
Z^{\prime i}=\mathcal{S}^{\prime i0}$. \emph{According to equation} (\ref{e}%
), \emph{a new \textquotedblleft time-space\textquotedblright\ 4D spin} $Z$
\emph{is introduced and it is a physical quantity in the same measure as it
is the usual \textquotedblleft space-space\textquotedblright\ 4D spin} $S$.
Both 4D vectors $S$\ and $Z$\ transform under the LT as any other 4D vector
transforms, i.e., the components in the standard basis transform like in
equation (\ref{el}). The 4D vector $S$\ transforms again to $S^{\prime }$\
and there is no mixing with $Z$. As already stated the transformations of
the 3-spin from equation (45) in [8] (equation (11.159) in [1]) are a
typical example of the AT and they have nothing to do with the
mathematically correct LT of the 4D intrinsic angular momentum vector $S$.

Hence, the correct introduction of the total angular momentum has to be
expressed in terms of the primary quantities as

\begin{equation}
J=M+\mathcal{S},  \label{jms}
\end{equation}%
or, in the tensor notation as $J^{ab}=M^{ab}+S^{ab}$, and not in the form of
equation (42) in [8]. Only in the case that $v=u$,\ i.e., the observer is
comoving with the particle, one could have $J_{s}^{a}=M_{s}^{a}+S^{a}$,
which stands instead of equation (42) in [8]. However, together with that
equation we have another equally important and physical equation $%
J_{t}^{a}=M_{t}^{a}+Z^{a}$.\ This is a fundamental difference between our
approach which exclusively deals with 4D GQs and the treatment from [8].

In [10] (earlier in [37]) a fundamental result is obtained by a consistent
application of the 4D GQs and the relations like (\ref{fm}) and (\ref{ebv}).
First, the generalized Uhlenbeck-Goudsmit hypothesis is formulated as the
relation which connects the dipole moment tensor $D^{ab}$ and the spin
four-tensor $S^{ab}$, $D^{ab}=g_{S}S^{ab}$, equation (9) in [10], instead of
the usual relation between the 3D vectors, the magnetic moment $\mathbf{m}$
and the spin 3D vector $\mathbf{S}$, $\mathbf{m}=\gamma _{S}\mathbf{S}$.
Then, both $D^{ab}$ and $S^{ab}$ are decomposed like in (\ref{fm}) into the
dipole moment 4-vectors $m^{a}$, $d^{a}$, equation (2) in [10], and the\
intrinsic angular momentum 4-vectors, the usual $S^{a}$ and the new one $%
Z^{a}$, equation (8) in [10], which is equation (\ref{e}) here. It is
obtained in a mathematically correct procedure that $d^{a}$, the electric
dipole moment of a fundamental particle, is determined by $Z^{a}$ and not,
as generally accepted, by the spin 3D vector $\mathbf{S}$. The connections
between the dipole moments $m^{a}$ and $d^{a}$ and the corresponding
intrinsic angular momentums $S^{a}$ and $Z^{a}$, respectively, are given by
equation (10) in [10]
\begin{equation}
m^{a}=cg_{S}S^{a},\ d^{a}=g_{S}Z^{a}.  \label{dm}
\end{equation}%
In the particle's rest frame and the $\{e_{\mu }^{\prime }\}$ basis, $%
u^{a}=ce_{0}^{\prime }$ and $d^{\prime 0}=m^{\prime 0}=0$, $d^{\prime
i}=g_{S}Z^{\prime i}$, $m^{\prime i}=cg_{S}S^{\prime i}$.

Furthermore, an important result is obtained in [38] by using the
mathematical theorem from section 3.3. In that paper, [38], we have reported
the relativistic generalizations of the usual commutation relations for the
components of the 3D orbital angular momentum $\mathbf{L}$. From the Lie
algebra of the Poincar\'{e} group we know that%
\begin{equation}
\lbrack M^{\mu \nu },M^{\rho \sigma }]=-i\text{%
h{\hskip-.2em}\llap{\protect\rule[1.1ex]{.325em}{.1ex}}{\hskip.2em}%
}(-g^{\nu \rho }M^{\mu \sigma }+g^{\mu \rho }M^{\nu \sigma }+g^{\mu \sigma
}M^{\nu \rho }-g^{\nu \sigma }M^{\mu \rho }).  \label{crm}
\end{equation}%
Taking into account the decomposition of the components $M^{\mu \nu }$, (\ref%
{lg}), into $M_{s}^{\mu }$ and $M_{t}^{\mu }$ (they are now operators),
where, for a macroscopic observer, $v^{\mu }$ can be taken as the classical
velocity of the observer (the components), i.e., not the operator. This
leads to the new commutation relations, equation (3) in [38],%
\begin{eqnarray}
\lbrack M_{s}^{\mu },M_{s}^{\nu }] &=&(i\hslash /c)\varepsilon ^{\mu \nu
\alpha \beta }M_{sa}v_{\beta },\ [M_{t}^{\mu },M_{t}^{\nu }]=(-i\hslash
/c)\varepsilon ^{\mu \nu \alpha \beta }M_{sa}v_{\beta },  \notag \\
\lbrack M_{s}^{\mu },M_{t}^{\nu }] &=&(i\hslash /c)\varepsilon ^{\mu \nu
\alpha \beta }M_{ta}v_{\beta },  \label{i}
\end{eqnarray}%
which, in the $\gamma _{0}$-frame, where $M_{s}^{0}=M_{t}^{0}=0$, reduce to
the usual commutators for the components of $\mathbf{L}$ and $\mathbf{K}$
(as operators), see, e.g., [39] equations (2.4.18) - (2.4.20). It is worth
noting that the same commutation relations (\ref{i}) can be obtained using $%
M_{s}^{\mu }$ and $M_{t}^{\mu }$ expressed in terms of $M^{\mu \nu }$,
equation (\ref{lg}), and the relativistic generalization of the fundamental
commutation relations, i.e., the worldspace fundamental commutation
relations, see, e.g., [40],
\begin{equation}
\lbrack x_{\mu },p_{\nu }]=i\text{%
h{\hskip-.2em}\llap{\protect\rule[1.1ex]{.325em}{.1ex}}{\hskip.2em}%
}\delta _{\mu \nu },\ [x_{\mu },x_{\nu }]=[p_{\mu },p_{\nu }]=0.  \label{ix}
\end{equation}

The same commutators as in (\ref{i}) have to hold for the intrinsic angular
momentums (the components) $S^{\mu }$ and $Z^{\mu }$; $S^{\mu }$ replaces $%
M_{s}^{\mu }$, $Z^{\mu }$ replaces $M_{t}^{\mu }$ and the velocity of the
particle (the components) $u^{\mu }$ replaces the velocity of the observer $%
v^{\mu }$, equation (4) in [38],%
\begin{eqnarray}
\lbrack S^{\mu },S^{\nu }] &=&(i\hslash /c)\varepsilon ^{\mu \nu \alpha
\beta }S_{a}u_{\beta },\ [Z^{\mu },Z^{\nu }]=(-i\hslash /c)\varepsilon ^{\mu
\nu \alpha \beta }S_{a}u_{\beta },  \notag \\
\lbrack S^{\mu },Z^{\nu }] &=&(i\hslash /c)\varepsilon ^{\mu \nu \alpha
\beta }Z_{a}u_{\beta }.  \label{sz}
\end{eqnarray}%
Usually, e.g., [41], only the commutators $[L_{i},L_{j}]$ and $[S_{i},S_{j}]$
appear.

Taking into account the relations (\ref{dm}), i.e., in components, $m^{\mu
}=cg_{S}S^{\mu }$, $d^{\mu }=g_{S}Z^{\mu }$ one can express the commutation
relations for $m^{\mu }$ and $d^{\mu }$ in terms of those for $S^{\mu }$ and
$Z^{\mu }$,%
\begin{equation}
\lbrack m^{\mu },m^{\nu }]=c^{2}g_{S}^{2}[S^{\mu },S^{\nu }],\ [d^{\mu
},d^{\nu }]=g_{S}^{2}[Z^{\mu },Z^{\nu }],\ [m^{\mu },d^{\nu
}]=cg_{S}^{2}[S^{\mu },Z^{\nu }].  \label{1}
\end{equation}%
what is equation (5) in [38].\bigskip \bigskip

\noindent \textbf{9. The electromagnetic field of a point charge in uniform
motion}\bigskip

\noindent It is worth mentioning that KS [8] and also the majority of
physicists consider that if the electric field would be transformed by the
LT again into the electric field as in (\ref{LTE}), i.e., as if in their
relation (35) $E^{\prime 0^{\prime }}$\ would be different from zero, then
it would imply, [8]: \textquotedblleft that moving electrons produce no
magnetic field.\textquotedblright\ In section 5.6 in [34] the
electromagnetic field of a point charge in uniform motion is treated in
detail. There it is shown that the formulation of that problem with the 4D
fields and their LT (\ref{aec}), (\ref{LTE}) is mathematically completely
correct but its physical interpretation is different than in the usual
formulation with the 3D fields and their AT. The above assertion from [8] is
caused by their \emph{incorrect} assumption that for both relatively moving
inertial observers $o$ and $o^{\prime }$ the temporal component of the
electric 4D vector is zero $E^{0}=E^{\prime 0}=0$, i.e., that $E^{\mu
}(o)=(0,\mathbf{E)}$ and $E^{\mu ^{\prime }}(o^{\prime })=(0,\mathbf{E}%
^{\prime }\mathbf{)}$. The consideration presented in 5.6.2 - 5.6.2.2 in
[34] explicitly shows that their assertion is not correct and that the
formulation with the 4D fields that transform according to the LT (\ref{aec}%
), (\ref{LTE}) simply explains the existence of the electric \emph{and
magnetic fields} for a moving electron.\bigskip \medskip

\noindent \textit{9.1. The bivector field }$F$\textit{\bigskip }

\noindent Here we shall briefly quote the main results from [34]. \emph{In
the 4D formulation the primary quantity is the the bivector field} $F$. The
expression for $F$ for an arbitrary motion of a point charge is given in
[14] by equations (10) (coordinate-free quantities) and (11) (CBGQs).
Particularly, for a charge $Q$ moving with constant 4D velocity vector $u$, $%
F$ is given by equation (12) in [14] (coordinate-free quantities), i.e.,
equation (65) in [34]

\begin{equation}
F(x)=G(x\wedge (u/c)),\quad G=kQ/\left\vert x\wedge (u/c)\right\vert ^{3},
\label{fcf}
\end{equation}%
where $k=1/4\pi \varepsilon _{0}$. $G$ is a number, a Lorentz scalar. The
geometric character of $F$ is contained in $x\wedge (u/c)$. If that $F$\ is
written as a CBGQ in the standard basis it is

\begin{equation}
F=(1/2)F^{\mu \nu }\gamma _{\mu }\wedge \gamma ;F^{\mu \nu }=G(1/c)(x^{\mu
}u^{\nu }-x^{\nu }u^{\mu }),G=kQ/[(x^{\mu }u_{\mu })^{2}-c^{2}x^{\mu }x_{\mu
}]^{3/2}.  \label{fmn}
\end{equation}%
In order to find the explicit expression for $F$ from (\ref{fmn}) in the $%
S^{\prime }$ frame in which the charge $Q$ is at rest one has simply to put
into (\ref{fmn})\ that $u=c\gamma _{0}^{\prime }$ with $\gamma _{0}^{\prime
\mu }=(1,0,0,0)$. Then, $F=(1/2)F^{\prime \mu \nu }\gamma _{\mu }^{\prime
}\wedge \gamma _{\nu }^{\prime }$ and

\begin{equation}
F=F^{\prime i0}(\gamma _{i}^{\prime }\wedge \gamma _{0}^{\prime
})=Gx^{\prime i}(\gamma _{i}^{\prime }\wedge \gamma _{0}^{\prime }),\quad
G=kQ/(x^{\prime i}x_{i}^{\prime })^{3/2}.  \label{q1}
\end{equation}%
In $S^{\prime }$ and in the standard basis, the basis components $F^{\prime
\mu \nu }$ of the bivector $F$ are obtained from (\ref{fmn}) and they are:
\begin{equation}
F^{\prime i0}=-F^{\prime 0i}=kQx^{\prime i}/(x^{\prime i}x_{i}^{\prime
})^{3/2},\quad F^{\prime ij}=0.  \label{fci}
\end{equation}%
In the charge's rest frame there are only components $F^{\prime i0}$, which
are the same as the usual components of the 3D electric field $\mathbf{E}$
for a charge at rest.

In the same way we find the expression for $F$ (\ref{fmn}) in the $S$ frame
in which the charge $Q$ is moving, i.e., $u=u^{\mu }\gamma _{\mu }$ with $%
u^{\mu }/c=(\gamma ,\gamma \beta ,0,0)$. Then
\begin{gather}
F=G\gamma \lbrack (x^{1}-\beta x^{0})(\gamma _{1}\wedge \gamma
_{0})+x^{2}(\gamma _{2}\wedge \gamma _{0})+x^{3}(\gamma _{3}\wedge \gamma
_{0})  \notag \\
-\beta x^{2}(\gamma _{1}\wedge \gamma _{2})-\beta x^{3}(\gamma _{1}\wedge
\gamma _{3})],\quad G=kQ/[\gamma ^{2}(x^{1}-\beta
x^{0})^{2}+(x^{2})^{2}+(x^{3})^{2}]^{3/2}.  \label{fsq}
\end{gather}%
In $S$ and in the standard basis, the basis components $F^{\mu \nu }$ of the
bivector $F$ are again obtained from (\ref{fmn}) and they are%
\begin{eqnarray}
F^{10} &=&G\gamma \lbrack (x^{1}-\beta x^{0}),\ F^{20}=G\gamma x^{2},\
F^{30}=G\gamma x^{3},  \notag \\
F^{21} &=&G\gamma \beta x^{2},\ F^{31}=G\gamma \beta x^{3},\ F^{32}=0.
\label{fi0}
\end{eqnarray}%
The expression for $F$ as a CBGQ in the $S$ frame can be find in another way
as well, i.e., to make the LT of the quantities from (\ref{q1}). Observe
that the CBGQs from (\ref{q1}) and (\ref{fsq}), which are the
representations of the bivector $F$ in $S^{\prime }$ and $S$ respectively,
are equal, $F$ from (\ref{q1}) $=$ $F$ from (\ref{fsq}); \emph{they are the
same quantity }$F$ \emph{from} (\ref{fcf}), i.e., (\ref{fmn}), \emph{for
observers in} $S^{\prime }$ \emph{and} $S$. It can be seen from (\ref{fi0})
that $F^{i0}$ \emph{and} $F^{ij}$ are different from zero for a moving
charge and they are the same as the usual components of the 3D fields $%
\mathbf{E}$ and $\mathbf{B}$, respectively. But, as already discussed and as
seen from (\ref{fc}) and (\ref{fmn}) only the whole $F$, which contains
components \emph{and the bivector basis}, is properly defined physical
quantity.\bigskip \medskip

\noindent \textit{9.2. The expressions for the 4D }$E$ \textit{and} $B$%
\bigskip

\noindent \textit{The general expressions\medskip }

\noindent From the known $F$ (\ref{fmn}) and\ the relations (\ref{ebv}) we
can construct in a mathematically correct way the 4D vectors $E$ and $B$ for
a charge $Q$ moving with constant velocity $u$. If written as CBGQs in the
standard basis they are given by equation (73) in [34]

\begin{eqnarray}
E &=&E^{\mu }\gamma _{\mu }=(G/c^{2})[(u^{\nu }v_{\nu })x^{\mu }-(x^{\nu
}v_{\nu })u^{\mu }]\gamma _{\mu },  \notag \\
B &=&B^{\mu }\gamma _{\mu }=(G/c^{3})\varepsilon ^{\mu \nu \alpha \beta
}x_{\nu }u_{\alpha }v_{\beta }\gamma _{\mu },  \label{ecb}
\end{eqnarray}%
where $G$\ is from (\ref{fmn}). \emph{The vectors} $E$ \emph{and} $B$ \emph{%
are explicitly} \emph{observer dependent}, i.e., dependent on $v$. For the
same $F$ the vectors $E$ and $B$ will have different expressions depending
on the velocity of observers who measure them. It is visible from (\ref{ecb}%
) that $E$ and $B$ depend on \emph{two} velocity 4D vectors $u$ and $v$,
whereas the usual 3D vectors $\mathbf{E}$ and $\mathbf{B}$ depend only on
the 3-velocity of the charge $Q$. Note also that although $E$ and $B$ as the
CBGQs from (\ref{ecb}) depend not only on $u$ but on $v$ as well the
electromagnetic field $F$ from (\ref{fmn}) does not contain the velocity of
the observer $v$. This result directly proves that \emph{the electromagnetic
field} $F$ \emph{is the primary quantity from which the observer dependent} $%
E$ \emph{and} $B$ \emph{are derived.} The expressions for $E$ and $B$ from (%
\ref{ecb}) correctly describe fields in all cases simply specifying $u$
\emph{and} $v$ and this assertion holds not only for the $\left\{ \gamma
_{\mu }\right\} $ basis but for the $\{r_{\mu }\}$ basis as well, i.e., the
relation like (\ref{erc}) holds for the expressions from (\ref{ecb}).
However, observe that, as already mentioned several times, \emph{the 4D
fields} $E$ \emph{and} $B$ \emph{and the usual 3D fields} $\mathbf{E}$ \emph{%
and} $\mathbf{B}$ \emph{have the same physical interpretation only in the} $%
\gamma _{0}$ - \emph{frame} \emph{with the} $\left\{ \gamma _{\mu }\right\} $
\emph{basis in which} $E^{0}=B^{0}=0$. In section 5.6.2.1 in [34] the
general expression (\ref{ecb}) for the 4D $E$ and $B$ is specified to the
case when the $\gamma _{0}$ - frame is the rest frame of the charge $Q$, the
$S^{\prime }$ frame, $v=c\gamma _{0}^{\prime }=u$, whereas in section
5.6.2.2 the same is made in the case when the $\gamma _{0}$ - frame is the
laboratory frame, the $S$ frame, $v=c\gamma _{0}$, in which the charge $Q$
is moving, $u^{\mu }=(\gamma c,\beta \gamma c,0,0)$.\bigskip

\noindent \textit{The} $\gamma _{0}$ - \textit{frame is the rest frame of
the charge} $Q$, \textit{the} $S^{\prime }$ \textit{frame\medskip }

\noindent If the $\gamma _{0}$ - frame is the $S^{\prime }$ frame, $%
v=c\gamma _{0}^{\prime }=u$, then (\ref{ecb}) yields that $B=0$ and only an
electric field (Coulomb field) remains, which is in agreement with the usual
3D formulation. Hence, it follows from (\ref{ecb}) that
\begin{equation}
E=E^{\prime i}\gamma _{i}^{\prime }=Gx^{\prime i}\gamma _{i}^{\prime },\quad
E^{\prime 0}=0,\quad G=kQ/(x^{\prime i}x_{i}^{\prime })^{3/2};\quad
B=B^{\prime \mu }\gamma _{\mu }^{\prime }=0.  \label{bf}
\end{equation}%
The components in (\ref{bf}) agree, as it is expected, with the usual result
with the 3D fields, e.g., with components in equation (11) in the first
paper in [32]. Now comes the essential difference relative to all usual
approaches. In order to find the representations of $E$ and $B$ in $S$,
i.e., the CBGQs $E^{\mu }\gamma _{\mu }$ and $B^{\mu }\gamma _{\mu }$, we
can either perform the LT of $E^{\prime \mu }\gamma _{\mu }^{\prime }$ and $%
B^{\prime \mu }\gamma _{\mu }^{\prime }$ that are given by (\ref{bf}), or
simply to take in (\ref{ecb}) that \emph{both} the charge $Q$ \emph{and the
\textquotedblleft fiducial\textquotedblright\ observers} are moving relative
to the observers in $S$; $v^{\mu }=u^{\mu }=(\gamma c,\beta \gamma c,0,0)$.
This yields equation (\ref{bs}) ((75) in [34]), i.e., the CBGQs $E^{\mu
}\gamma _{\mu }$ and $B^{\mu }\gamma _{\mu }$ in $S$ with the condition that
the \textquotedblleft fiducial\textquotedblright\ observers are in $%
S^{\prime }$, $v=c\gamma _{0}^{\prime }$, which is the rest frame of the
charge $Q$, $u=c\gamma _{0}^{\prime }$,

\begin{gather}
E=E^{\mu }\gamma _{\mu }=G[\beta \gamma ^{2}(x^{1}-\beta x^{0})\gamma
_{0}+\gamma ^{2}(x^{1}-\beta x^{0})\gamma _{1}+  \notag \\
x^{2}\gamma _{2}+x^{3}\gamma _{3}],\quad B=B^{\mu }\gamma _{\mu }=0,
\label{bs}
\end{gather}%
where $G$ is that one from (\ref{fsq}). The result (\ref{bs}) significantly
differs from the result obtained by the AT, equations (12a), (12b) in [32].
Under the LT the electric field vector transforms again to the electric
field vector and the same for the magnetic field vector. It is worth
mentioning that, in contrast to the conventional results, it holds that $%
E^{\prime \mu }\gamma _{\mu }^{\prime }$ from (\ref{bf}) is $=E^{\mu }\gamma
_{\mu }$ from (75) in [34]; \emph{they are the same quantity} $E$ \emph{for
all relatively moving inertial observers.} The same holds for $B$, $%
B^{\prime \mu }\gamma _{\mu }^{\prime }$ from (\ref{bf}) is $=B^{\mu }\gamma
_{\mu }$ from (\ref{bs}) and they are $=0$ for all observers. Furthermore,
observe that in $S^{\prime }$ there are only the spatial components $%
E^{\prime i}$, whereas in $S$, as seen from (\ref{bs}), there is also the
temporal component $E^{0}$ as a consequence of the LT.\bigskip

\noindent \textit{The} $\gamma _{0}$ - \textit{frame is the} \textit{%
laboratory frame, the} $S$ \textit{frame\medskip }

\noindent Now, let us take that the \textquotedblleft
fiducial\textquotedblright\ observers are in $S$, $v=c\gamma _{0}$, in which
the charge $Q$ is moving, $u^{\mu }=(\gamma c,\beta \gamma c,0,0)$. In
contrast to the previous case, \emph{both} $E$ \emph{and} $B$ are different
from zero. The expressions for the CBGQs $E^{\mu }\gamma _{\mu }$ and $%
B^{\mu }\gamma _{\mu }$ in $S$ can be simply obtained from (\ref{ecb})
taking in it that $v=c\gamma _{0}$ and $u^{\mu }=\gamma c\gamma _{0}+\beta
\gamma c\gamma _{1}$. This yields that $E^{0}=B^{0}=0$ (from $v=c\gamma _{0}$%
) and the spatial parts are
\begin{eqnarray}
E &=&E^{i}\gamma _{i}=G\gamma \lbrack (x^{1}-\beta x^{0})\gamma
_{1}+x^{2}\gamma _{2}+x^{3}\gamma _{3}],  \notag \\
B &=&B^{i}\gamma _{i}=(G/c)[0\gamma _{1}-\beta \gamma x^{3}\gamma _{2}+\beta
\gamma x^{2}\gamma _{3}],  \label{seb}
\end{eqnarray}%
where $G$ is again as in (\ref{fsq}). The 4D vector fields $E$ and $B$ from (%
\ref{seb}) can be compared with the usual expressions for the 3D fields $%
\mathbf{E}$ and $\mathbf{B}$ of an uniformly moving charge, e.g., from
equations (12a), (12b) in [32]. It is visible that they are similar, but $E$
and $B$ in (\ref{seb}) are the 4D fields and all quantities in (\ref{seb})
are correctly defined in the 4D spacetime, which transform by the LT,
whereas the fields in equations (12a), (12b) in [32] are the 3D fields that
transform according to the AT.

In order to find the representations of $E$ and $B$ in $S^{\prime }$, i.e.,
the CBGQs $E^{\prime \mu }\gamma _{\mu }^{\prime }$ and $B^{\prime \mu
}\gamma _{\mu }^{\prime }$, we can either perform the LT of $E^{\mu }\gamma
_{\mu }$ and $B^{\mu }\gamma _{\mu }$ that are given by (\ref{seb}), or
simply to take in (\ref{ecb}) that relative to $S^{\prime }$ the
\textquotedblleft fiducial\textquotedblright\ observers are moving with $%
v=v^{\prime \mu }\gamma _{\mu }^{\prime }$, $v^{\prime \mu }=(c\gamma
,-\beta \gamma c,0,0)$, and the charge $Q$ is at rest relative to the
observers in $S^{\prime }$, $u^{\prime \mu }=(c,0,0,0)$. This yields the
CBGQs $E^{\prime \mu }\gamma _{\mu }^{\prime }$ and $B^{\prime \mu }\gamma
_{\mu }^{\prime }$ in $S^{\prime }$ with the condition that the
\textquotedblleft fiducial\textquotedblright\ observers are in $S$, $%
v=c\gamma _{0}$,%
\begin{eqnarray}
E &=&E^{\prime \mu }\gamma _{\mu }^{\prime }=G\gamma \lbrack -\beta
x^{\prime 1}\gamma _{0}^{\prime }+x^{\prime 1}\gamma _{1}^{\prime
}+x^{\prime 2}\gamma _{2}^{\prime }+x^{\prime 3}\gamma _{3}^{\prime }],
\notag \\
B &=&B^{\prime \mu }\gamma _{\mu }^{\prime }=(G/c)[0\gamma _{0}^{\prime
}+0\gamma _{1}^{\prime }-\beta \gamma x^{\prime 3}\gamma _{2}^{\prime
}+\beta \gamma x^{\prime 2}\gamma _{3}^{\prime }],  \label{sc}
\end{eqnarray}%
where $G$ is as in (\ref{bf}). Again, as in the case that $v=c\gamma
_{0}^{\prime }$, it holds that $E^{\mu }\gamma _{\mu }$ from (\ref{seb}) is $%
=E^{\prime \mu }\gamma _{\mu }^{\prime }$ from (\ref{sc}); they are the same
quantity $E$ for all relatively moving inertial observers. The same holds
for $B^{\mu }\gamma _{\mu }$ from (\ref{seb}) which is $=B^{\prime \mu
}\gamma _{\mu }^{\prime }$ from (\ref{sc}) and \emph{they are both different
from zero}. Note that in this case there are only the spatial components $%
E^{i}$ in $S$, whereas in $S^{\prime }$ there is also the temporal component
$E^{\prime 0}$ as a consequence of the LT.

It is visible from (\ref{sc}) that if the $\gamma _{0}$ - frame is the lab
frame ($v=c\gamma _{0}$) in which the charge $Q$ is moving then $E^{\prime
\mu }\gamma _{\mu }^{\prime }$ and $B^{\prime \mu }\gamma _{\mu }^{\prime }$
in the rest frame of the charge $Q$, the $S^{\prime }$ frame, are completely
different than those from (\ref{bf}); in (\ref{sc}) $B^{\prime \mu }\gamma
_{\mu }^{\prime }$ is different from zero and the representation of $E$\
contains also the term $E^{\prime 0}\gamma _{0}^{\prime }$.

It has to be emphasized that \emph{all four expressions for} $E$ \emph{and} $%
B$, (\ref{bf}), (\ref{bs}), (\ref{seb}) \emph{and} (\ref{sc}), \emph{are the
special cases of} $E$ \emph{and} $B$ \emph{given by} (\ref{ecb}). \emph{They
all give the same} $F$ \emph{from} (\ref{fmn}), \emph{which is the
representation (CBGQ) of} $F$ \emph{given by the basis free, abstract,
bivector} (\ref{fcf}).\bigskip \bigskip

\noindent \textbf{10. Comparison with the experiments}\bigskip

\noindent The approach with 4D GQs and their mathematically correct LT is in
a true agreement, independent of the chosen inertial reference frame and of
the chosen basis in it, with experiments in electromagnetism. This is
already explicitly shown in [22, 12] for the motional emf, in [23] for the
Faraday disk and in [33, 14] for the Trouton-Noble experiment.

A nice example that illustrates the fundamental difference between the LT
like (\ref{aec}), (\ref{LTE}) and the AT (\ref{ee}), i.e., between the
approach with 4D GQs and the usual approach with the 3D vectors is presented
in the discussion of the motional electromotive force (emf) in sections 5 -
5.2 in [22].

In section 5.1 in [22] the motional emf $\varepsilon $ is calculated using
the 3D Lorentz force, $\mathbf{F}_{L}\mathbf{=}q\mathbf{E}+q\mathbf{U}\times
\mathbf{B}$, and the AT for the 3D $\mathbf{E}$ and $\mathbf{B}$, equation
(11.149) in [1]. The emf $\varepsilon $ of a complete circuit is defined by
means of $\mathbf{F}_{L}$ that acts on a charge $q$, which is at rest
relative to the section $\mathbf{dl}$ of the circuit

\begin{equation}
\varepsilon =\oint (\mathbf{F}_{L}\mathbf{/}q)\cdot \mathbf{dl},  \label{eps}
\end{equation}%
equation (26) in [22]. Observe that it is implicitly assumed in (\ref{eps})
that the integral is taken over the whole circuit at the same moment of time
in $S$, say $t=0$. Then it is assumed that in the laboratory frame $S$ a
conducting bar is moving in a steady uniform magnetic field (3D vector) $%
\mathbf{B=-}B\mathbf{k}$ with velocity 3D vector $\mathbf{U}$ parallel to
the $x$ axis. The length of the bar is $l$ and it moves parallel to the $y$
axis. There is no external applied electric field in $S$, $\mathbf{E=}0$ and
the components of $\mathbf{B}$ are $\mathbf{(}0,0\mathbf{,-}B\mathbf{)}$,
which yields that the emf $\varepsilon $ is
\begin{equation}
\varepsilon =\int_{o}^{l}UBdy=UBl,  \label{uef}
\end{equation}%
equation (27) in [22]. Note that in $S$ the emf $\varepsilon $ is determined
only by \emph{the contribution of the magnetic part of} the 3D Lorentz force
$\mathbf{F}_{L}$, i.e., $q\mathbf{U}\times \mathbf{B}$. On the other hand,
in $S^{\prime }$ the conducting bar is at rest. The usual explanation is of
this kind. If in $S$ $\mathbf{E=}0$ and the components of $\mathbf{B}$ are $%
\mathbf{(}0,0\mathbf{,-}B\mathbf{)}$\ then, \emph{according to the AT }of
the 3D $\mathbf{E}$ and $\mathbf{B}$, equation (11.148) in [1], the observer
in the $S^{\prime }$ frame `sees' $E_{y}^{\prime }=\gamma UB$ and $%
B_{z}^{\prime }=-\gamma B$. Hence \emph{in} $S^{\prime }$ there is not only
the magnetic field but \emph{an induced electric field} as well. The
calculation of $\varepsilon ^{\prime }$ in\emph{\ }$S^{\prime }$ yields that
the contribution of $B_{z}^{\prime }$ to the emf $\varepsilon ^{\prime }$ is
zero and \emph{only the contribution of} $E_{y}^{\prime }$ \emph{remains},
which is
\begin{equation}
\varepsilon ^{\prime }=\int_{o}^{l}\gamma UBdy=\gamma UBl,  \label{uec}
\end{equation}%
equation (29) in [22]. Observe that the integral in (\ref{uec}) is again
taken at the same moment of time but now $t^{\prime }$ in $S^{\prime }$,
which can be arbitrarily chosen, say $t^{\prime }=0$, or $t^{\prime }=10s$,
... . \emph{The moments of time} $t$ \emph{in} $S$ \emph{and} $t^{\prime }$
\emph{in} $S^{\prime }$ \emph{are not connected in any way.} The LT cannot
transform the moment of time $t$ in $S$ again, exclusively, to some $%
t^{\prime }$ in $S^{\prime }$. According to the LT, to one $t$ in $S$ will
correspond many $t^{\prime }$ in $S^{\prime }$ depending on the spatial
position in $S^{\prime }$; $t=\gamma (t^{\prime }+Ux^{\prime }/c^{2})$. This
remark clearly shows that the usual definition of $\varepsilon $, (\ref{eps}%
), is not relativistically correct definition.

It is visible that \emph{the emf }$\varepsilon ^{\prime }$ \emph{in }$%
S^{\prime }$ \emph{is not equal to the emf} $\varepsilon $ \emph{determined}
\emph{in} $S$.

\begin{equation}
\varepsilon =UBl,\ \varepsilon ^{\prime }=\gamma UBl,\quad \varepsilon
^{\prime }\neq \varepsilon .  \label{eec}
\end{equation}%
$\varepsilon ^{\prime }$ is not much different from $\varepsilon $ only if $%
U\ll c$, i.e., $\gamma \simeq 1$. This means \emph{the principle of
relativity is not satisfied}; \emph{the emf obtained by the application of
the AT for the }3D $\mathbf{E}$ \emph{and} $\mathbf{B}$ \emph{is different
for relatively moving 4D observers.} That result explicitly shows that the
AT of the 3D $\mathbf{E}$ and $\mathbf{B}$ are not the correct relativistic
transformations, i.e., they are not the LT. Thus, it is not true that the
conventional formalism correctly describes even such simple experiment. This
is a simple but \emph{completely correct calculation,} which reveals a
fundamental flaw in the usual formulations with the 3D $\mathbf{E}$ and $%
\mathbf{B}$ and their AT. The fact that $\varepsilon $ and $\varepsilon
^{\prime }$ do not significantly differ for low velocities is completely
irrelevant; \emph{the principle of relativity is not satisfied }in the usual
approach.

On the other hand, in section 5.2 in [22] the emf $\varepsilon $ is
calculated using the 4D GQs. The Lorentz force $K_{L}$ is defined by
equations (\ref{lfa}) or (\ref{lf}). These expressions reveal the
fundamental difference between $K_{L}$ and the 3D Lorentz force $\mathbf{F}%
_{L}$; $K_{L}$ contains not only the 4-velocity $u$ of a charge $q$\ but
also the 4-velocity $v$ of the observer who measures 4D fields. \emph{Then
the emf} $\varepsilon $ \emph{is defined by equation }(35)\emph{\ in }[22]
\emph{as an invariant 4D quantity, the Lorentz scalar},
\begin{equation}
\varepsilon =\int_{\Gamma }(K_{L}/q)\cdot dl,  \label{emf4}
\end{equation}%
where \emph{vector} $dl$ is the infinitesimal \emph{spacetime length} and $%
\Gamma $ \emph{is the spacetime curve}. In the laboratory frame $S$ as the $%
\gamma _{0}$- frame, the observers are at rest $v=c\gamma _{0}$, whereas the
conducting bar is moving with velocity vector $u$, $u^{\mu }=(\gamma
c,\gamma U,0,0)$. Furthermore, $E=0$ and $B^{\mu }=(0,0,0,-B)$. Hence, $%
K_{L}^{0}=K_{L}^{1}=K_{L}^{3}=0$, but $K_{L}^{2}=\gamma qUB$, yielding that $%
\varepsilon =\int_{0}^{l}\gamma UBdy=\gamma UBl$ (equation (36) in [22]);
the emf $\varepsilon $ is determined by the contribution of the magnetic
part of $K_{L}$, i.e., $(q/c)\left[ (IB)\cdot v\right] \cdot u$.

Now comes the main difference relative to the usual approaches with the 3D
quantities. \emph{The expression for} $\varepsilon $ (\ref{emf4}) \emph{is
independent of the chosen reference frame and of the chosen basis in it.}
Hence, $\varepsilon $ \emph{is the same in} $S$ \emph{and in the relatively
moving} $S^{\prime }$ \emph{frame};
\begin{equation}
\varepsilon =\int_{\Gamma }(K_{L}^{\mu }/q)dl_{\mu }=\int_{\Gamma
}(K_{L}^{\prime \mu }/q)dl_{\mu }^{\prime }=\gamma UBl,  \label{ef5}
\end{equation}%
(equation (37) in [22]) and the same holds if the $\{r_{\mu }\}$ basis is
used. This means that \emph{the observers in} $S$ \emph{and} $S^{\prime }$
\emph{are `looking' at the same physical quantity} $\varepsilon $\ defined
by (\ref{emf4}).

Obviously, in contrast to the usual approaches, \emph{the principle of
relativity is naturally satisfied} \emph{in the approach with 4D GQs and
their mathematically correct LT}, like (\ref{LTE}). This result (\ref{ef5})
for $\varepsilon $ can be checked directly performing the LT of all vectors
from $S$ to $S^{\prime }$ as in [22]. In $S^{\prime }$ the 3-velocity $%
\mathbf{U}$ of a charge $q$ is zero, but the velocity vector $u$ \emph{is not%
}, $u=c\gamma _{0}$. From the viewpoint of the observers in $S^{\prime }$
the velocity vector $v$ of the \textquotedblleft fiducial\textquotedblright\
observers contains not only the temporal component as in $S$ ($v=c\gamma
_{0} $), but also the spatial component, $v^{\prime \mu }=(\gamma c,-\gamma
U,0,0) $. \emph{According to the LT,} like (\ref{LTE}), \emph{there is no
mixing of components of vectors of the electric and magnetic fields. This
means that in} $S^{\prime }$, \emph{as in} $S$, \emph{there is no electric
field!!}

In this particular case the LT yield that the components $B^{\prime \mu }$
in $S^{\prime }$ are the same as $B^{\mu }$ in $S$, $B^{\prime \mu }=B^{\mu
}=(0,0,0,-B)$, and the same holds for the components of the Lorentz force, $%
K_{L}^{\prime 0}=K_{L}^{\prime 1}=K_{L}^{\prime 3}=0$ and $K_{L}^{\prime
2}=K_{L}^{2}=\gamma qUB$. \emph{In }$S^{\prime }$, \emph{as in} $S$, \emph{%
there is only the magnetic part of the Lorentz force} \emph{and again only
that part determines the emf} $\varepsilon $, equation (37) in [22]. Note
that in this calculation all quantities are invariant under the passive LT,
e.g., $B=B^{\nu }\gamma _{\nu }=B^{\prime \nu }\gamma _{\nu }^{\prime }$, $%
v=v^{\nu }\gamma _{\nu }=v^{\prime \nu }\gamma _{\nu }^{\prime }$, $K=K^{\nu
}\gamma _{\nu }=K^{\prime \nu }\gamma _{\nu }^{\prime }$, etc., and the same
holds if the $\{r_{\mu }\}$ basis is used.

The same result as in section 5.2 in [22] is obtained in [12] but
exclusively dealing with $F$ and not with its decompositions (\ref{E2}) and (%
\ref{fm}).

The result that the conventional theory with the 3D $\mathbf{E}$ and $%
\mathbf{B}$ and their AT, equations (11.148) and (11.149) in [1], i.e., here
(\ref{ee}), yields different values for the motional emf $\varepsilon $ for
relatively moving inertial observers, $\varepsilon =UBl$ in $S$ and $%
\varepsilon ^{\prime }=\gamma UBl$ in $S^{\prime }$, equation (\ref{eec}),
whereas the approach with 4D GQs and their LT, e.g., (\ref{LTE}), yields
always the same value for $\varepsilon $, $\varepsilon =\gamma UBl$,
equation (\ref{ef5}), is very strong evidence that the usual approach is not
relativistically correct. \emph{It is for the experimentalists to find the
way to measure the emf} $\varepsilon $ \emph{with a great precision in order
to see that in the laboratory frame} $\varepsilon =\gamma UBl$ \emph{and not
simply} $\varepsilon =UBl$. Such an experiment would be a crucial experiment
that could verify from the experimental viewpoint the validity of the
formulation of the electromagnetism with the 4D GQs and their mathematically
correct LT, like (\ref{aec}), (\ref{LTE}).

Completely the same conclusions about the fundamental difference between the
conventional theory with the 3D $\mathbf{E}$ and $\mathbf{B}$ and the theory
with 4D GQs are obtained in [23], where an important experiment, the Faraday
disk, is considered in detail. Particularly important and instructive
comparison with experiments is the comparison with the Trouton-Noble
experiment that is presented [33]. That comparison is also given in the
formulation with $F$ in section 4 in [12]. In these papers, it is shown that
in the treatment with 4D GQs the Trouton-Noble paradox does not appear. The
presented explanations are in a complete agreement with the principle of
relativity and with the Trouton-Noble experiment without the introduction of
any additional torque, which must be necessarily introduced in all usual
approaches with the 3D quantities.

Furthermore, in [13], the constitutive relations and the magnetoelectric
effect in moving media are explained in a completely new way using 4D GQs.
In equation (17) in [13] it is shown how the polarization vector $P(x)$
depends on $E$, $B$, $u$, the bulk velocity vector of the medium and $v$,
the velocity vector of the observer who measures fields%
\begin{equation}
P^{\mu }\gamma _{\mu }=(\varepsilon _{0}\chi _{E}/c)[(1/c)(E^{\mu }v^{\nu
}-E^{\nu }v^{\mu })+\varepsilon ^{\mu \nu \alpha \beta }v_{\alpha }B_{\beta
}]u_{\nu }\gamma _{\mu },  \label{po}
\end{equation}%
whereas in equation (18) in [13] the same is shown for the magnetization
vector $M(x)$,%
\begin{equation}
M^{\mu }\gamma _{\mu }=\varepsilon _{0}\chi _{B}[(B^{\mu }v^{\nu }-B^{\nu
}v^{\mu })+(1/c)\varepsilon ^{\mu \nu \alpha \beta }E_{\alpha }v_{\beta
}]u_{\nu }\gamma _{\mu }.  \label{ma}
\end{equation}%
Both equations are written with CBGQs, whereas the corresponding equations
with AQs are equations (13) and (14) in [13]. In this geometric approach,
the relations (\ref{po}) and (\ref{ma}) replace the constitutive relations
with the 3D vectors, equations (23) and (24) that are derived in [13] and
which are equivalent to Minkowski's constitutive relations given by equation
(22) in [13]. The equations (13) and (14) in [13] are derived from the basic
constitutive relations for moving media, equations (11) and (12) in [13],
which are written in terms of the primary quantities for the electric and
magnetic fields, i.e., the electromagnetic field bivector $F$, the primary
quantity for the the polarization and magnetization, i.e., the generalized
magnetization-polarization bivector $\mathcal{M}$ and the electric and
magnetic susceptibility $\chi _{E}$, $\chi _{B}$,%
\begin{equation}
\mathcal{M}\cdot u=\varepsilon _{0}\chi _{E}F\cdot u,  \label{cr3}
\end{equation}

\begin{equation}
(I\mathcal{M})\cdot u=(\chi _{B}/\mu _{0}c^{2})u\cdot (IF)  \label{cr2}
\end{equation}%
Then, the decompositions of $F$ (\ref{E2}) and the similar one for $\mathcal{%
M}$ (\ref{M1}) are used to derive equations (13) and (14) in [13]. If
equations (13) and (14) in [13] with AQs are written in terms of CBGQs in
the standard basis then equations (\ref{po}) and (\ref{ma}) are obtained.
The last term in (\ref{po}) and that one in (\ref{ma}) describe the
magnetoelectric effect in a moving dielectric. The last term in (\ref{po})
shows that a moving dielectric becomes electrically polarized if it is
placed in a magnetic field, the Wilsons' experiment [39]. Let us take that
the laboratory frame, the $S$ frame, is the $\gamma _{0}$-frame ($v=c\gamma
_{0}$) in which the material medium, the $S^{\prime }$ frame, is moving with
velocity $u$. If in equation (\ref{po}) it is chosen that $E^{\mu
}=(0,0,0,0) $, $B^{\mu }=(0,0,0,-B^{3})$, $u^{\mu }=(\gamma _{u}c,\gamma
_{u}U^{1},0,0)$, then, in $S$, equation (\ref{po}) becomes equation (20) in
[13],%
\begin{equation}
P^{\mu }=(0,0,P^{2}=\varepsilon _{0}\chi _{E}\gamma _{u}U^{1}B^{3},0).
\label{pwi}
\end{equation}%
The components in (\ref{pwi}) correspond to the \textquotedblleft
translational\textquotedblright\ version of Wilsons' experiment [42].
Similarly, the last term in (\ref{ma}) shows that a moving dielectric
becomes magnetized if it is placed in an electric field, R\"{o}ntgen's
experiment [43]. If in equation (\ref{ma}) it is chosen that $B^{\mu
}=(0,0,0,0)$, $E^{\mu }=(0,0,-E^{2},0)$, $u^{\mu }=(\gamma _{u}c,\gamma
_{u}U^{1},0,0)$, then, in $S$, equation (\ref{ma}) becomes
\begin{equation}
M^{\mu }=(0,0,0,M^{3}=(\chi _{B}/\mu _{0}c^{2})\gamma _{u}U^{1}E^{2}).
\label{mrt}
\end{equation}%
The components in (\ref{mrt}) correspond to the \textquotedblleft
translational\textquotedblright\ version of R\"{o}ntgen's experiment [43].
Observe that in this geometric approach all quantities are correctly defined
4D quantity that correctly transform under the LT. The term in (\ref{po})
and (\ref{ma}) that describes the magnetoelectric effect is obtained without
any transformations by the correct mathematical procedure from the
fundamental constitutive relations (\ref{cr3}) and (\ref{cr2}). It is not so
in all previous approaches, e.g., [44], in which the 3D $\mathbf{E}$, $%
\mathbf{B}$, $\mathbf{P}$, $\mathbf{M}$,$\ \mathbf{D}$, $\mathbf{H}$, etc.
and their AT are used considering them as that they are the mathematically
correct LT. In sections 5.1 and 5.2 the constitutive relations with 4D GQs,
the relations (\ref{po}) and (\ref{ma}) here, are compared with Minkowski's
constitutive relations with the 3D vectors, i.e., with the equivalent
relations (23) - (25) in [13]. It is shown that there are important
differences between them, which could be experimentally examined.\bigskip
\bigskip

\noindent \textbf{11. Discussion and Conclusions}\bigskip

\noindent The main point in the whole paper is explicitly expressed by the
motto at the beginning of the text. In the 4D spacetime physical laws are
geometric, coordinate-free relationships between the 4D geometric,
coordinate-free quantities. This point of view is also adopted in the nice
textbook [7] but not in the consistent way. They still introduce the 3D
vectors and their transformations, e.g., in section 1.10 in [7] and this is
discussed in section 6 here. Similarly happens in [8], which is discussed in
section 7 here. A fully consistent application of this viewpoint is adopted
in Oziewicz's papers, see, e.g., [11]. The same viewpoint is adopted in all
my papers given in the references and including the present paper. Here, in
this paper, the mathematically correct proofs are given that the electric
and magnetic fields are properly defined vectors on the 4D spacetime,
sections 3.1 and 3.3. According to Oziewicz's proof from section 3.1, e.g., $%
\mathbf{E(r,}t\mathbf{)}$ (written in the usual notation) must have four
components (some of them can be zero) since it is defined on the 4D
spacetime and not, as usually considered, only three components. In section
3.3 it is taken into account that, as proved in [14], the primary quantity
for the whole electromagnetism is the electromagnetic field bivector $F$.\
The decomposition of $F$\ given by equation (\ref{E2}) expresses $F$\ in
terms of observer dependent electric and magnetic 4D vectors $E$ and $B$,
which are given by equation (\ref{E1}).\ Both equations (\ref{E2}) and (\ref%
{E1}) are with the abstract, coordinate-free quantities. This is in a sharp
contrast with the usual covariant approaches, e.g., [1, 4, 36] in which it
is considered that $F^{\alpha \beta }$ (the components implicitly taken in
the standard basis) is physically well-defined quantity. Moreover, these
components are considered to be six indepent components of the 3D $\mathbf{E}
$ and $\mathbf{B}$, see equations (\ref{ieb}) and (\ref{eb2}). Then, as
described in section 1, in these approaches [1, 4, 36],\ \emph{the
transformations of the components of} $\mathbf{E}$ \emph{and} $\mathbf{B}$ (%
\ref{ee}) \emph{are obtained supposing that they transform under the LT as
the components of} $F^{\alpha \beta }$ \emph{transform}, equation (\ref{fe}%
). The objections to such treatment are given in section 1, the objections
1), 2) and section 2.1, the objections 3), 4) and 5). From the mathematical
viewpoint all these objections are well-founded since they are based on the
following facts: 1) The bivector $F(x)$, as described in detail in [14] and
very briefly in section 3.2 here, is determined, for the given sources, by
the solutions of the equation (\ref{fef}), i.e., (\ref{mf}) (with CBGQs in
the $\left\{ \gamma _{\mu }\right\} $ basis) and not by the components of
the 3D $\mathbf{E}$ and $\mathbf{B}$. It is a 4D GQ and not only components.
It yields a complete description of the electromagnetic field without the
need for the introduction either the field vectors or the potentials. 2) As
seen from section 2 and particularly from equations (\ref{are}) and (\ref%
{FEr}) the identification of the components of the 3D $\mathbf{E}$ and $%
\mathbf{B}$ with the components of $F^{\alpha \beta }$ is synchronization
dependent. Moreover, it is completely meaningless in the \textquotedblleft
r\textquotedblright\ synchronization, i.e., in the $\left\{ r_{\mu }\right\}
$ basis. Both bases, the commonly used standard basis with Einstein's
synchronization and the $\left\{ r_{\mu }\right\} $ basis with the
\textquotedblleft r\textquotedblright\ synchronization are equally well
physical and relativistically correct bases.

Furthermore, it is proved in section 4.1 with the coordinate-free quantities
and the active LT and in section 4.2 with CBGQs and the passive LT that the
mathematically correct LT of, e.g., the electric field vector are given by (%
\ref{aec}) - (\ref{el}) and not by the AT of the 3D vectors equations
(11.148) and (11.149) in [1], i.e., equation (\ref{ee}) or equation (\ref{ut}%
) here.

In section 5.1 the same fundamental difference between the correct LT and
the usual AT of the 3D vectors is explicitly exposed using matrices. The
equations (\ref{ecm}) - (\ref{an}) refer to the correct LT of the components
in the standard basis of the electric field 4D vector in which the
transformed components $E^{\prime \mu }$\ are obtained as $E^{\prime \mu
}=c^{-1}F^{\prime \mu \nu }v_{\nu }^{\prime }$, i.e., \emph{both} $F^{\mu
\nu }$\emph{\ and the velocity of the observer }$v=c\gamma _{0}$\emph{\ are
transformed}$\ $by the matrix of the LT $A_{\nu }^{\mu }$ (the boost in the
direction $x^{1}$). It is visible from equation (\ref{an}) that the same
components are obtained as $E^{\prime \mu }=A_{\nu }^{\mu }E^{\nu }$\ and
they are the same as in (\ref{LTE}). This means that \emph{under the
mathematically correct LT the electric field 4D vector transforms again only
to the electric field 4D vector as any other 4D vector transforms.} As
stated at the end of section 5.1 if $E$\ is written as a CBGQ then again
holds the relation (\ref{erc}) as for any other CBGQ. On the other hand
equation (\ref{Em4}) refers to the AT in which the transformed components $%
E_{F}^{\prime \mu }$\ are obtained as $E_{F}^{\prime \mu }=c^{-1}F^{\prime
\mu \nu }v_{\nu }$, i.e., only $F^{\mu \nu }$ \emph{is transformed by the LT
but not the velocity of the observer} $v=c\gamma _{0}$. These transformed
components $E_{F}^{\prime \mu }$\ are the same as in equation (\ref{ut}).
The transformed spatial components $E_{F}^{\prime i}$\ are the same as are
the transformed components of the usual 3D vector $\mathbf{E}$, i.e., as in
equation (11.148) in [1].\ However, according to these transformations the
4D vector with $E^{0}=0$ is transformed in such a way that the transformed
temporal component is again zero, $E_{F}^{\prime 0}=0$. Hence, as stated in
section 5.1, such transformations cannot be the mathematically correct LT.

It can be concluded from the whole consideration in this paper that in the
4D spacetime an independent physical reality has to be attributed to the 4D
geometric quantities, coordinate-free quantities or the CBGQs, e.g., the
electromagnetic field bivector $F$, the 4D vectors of the electric $E$\ and
magnetic $B$ fields, etc., and not to the usual 3D quantities, e. g., the 3D
$\mathbf{E}$ and $\mathbf{B}$. This is the answer to the question what is
the nature of the electric and magnetic fields. Furthermore, the
mathematically correct LT are properly defined on the 4D spacetime. They can
correctly transform only the 4D quantities like $E$\ and $B$, the
transformations (\ref{aec}) - (\ref{el}), according to which, e.g., \emph{%
the electric field 4D vector transforms again only to the electric field 4D
vector as any other 4D vector transforms.} The LT cannot act on the 3D
quantities like the 3D $\mathbf{E}$ and $\mathbf{B}$, which means that the
usual transformations of the 3D quantities, e.g., the 3D vectors $\mathbf{E}$
and $\mathbf{B}$, equations (11.148) and (11.149) in [1], i.e., equation (%
\ref{ee}) or equation (\ref{ut}) here, are not the LT, but the
mathematically incorrect transformations in the 4D spacetime, i.e., the AT.
This is the answer to the question how the fields transform.\bigskip
\bigskip

\noindent \textbf{References\bigskip }

\noindent \lbrack 1] Jackson J D 1998 \textit{Classical Electrodynamics} 3rd
edn (New York: Wiley)

\noindent \lbrack 2] Ivezi\'{c} T 2013 \textit{J. Phys.: Conf. Ser.} \textbf{%
437} 012014

\noindent \lbrack 3] Ivezi\'{c} T 2001 \textit{Found. Phys.} \textbf{31} 1139

Ivezi\'{c} T 2002 Annales Fond. \textbf{27} 287

\noindent \lbrack 4] Griffiths D J 2013 \textit{Introduction to
Electrodynamics} 4th edn (Pearson)

\noindent \lbrack 5] Ivezi\'{c} T 2002 \textit{Found. Phys. Lett.} \textbf{15%
} 27

Ivezi\'{c} T 2001 arXiv: physics/0103026

Ivezi\'{c} T 2001 arXiv: physics/0101091

\noindent \lbrack 6] Rohrlich F 1966 \textit{Nuovo Cimento B} \textbf{45} 76

\noindent \lbrack 7] Blandford R D and Thorne K S 2002-2003 \textit{%
Applications of classical }

\textit{physics} (California Institute of Technology)

\noindent \lbrack 8] Klajn B and Smoli\'{c} I 2013 \textit{Eur. J. Phys. }%
\textbf{34} 887

\noindent \lbrack 9] Einstein A 1905 \textit{Ann. Physik.} \textbf{17} 891

Perrett W and Jeffery G B 1952 \textit{The Principle of Relativity}

(New York: Dover) (Engl.Transl.)

\noindent \lbrack 10] Ivezi\'{c} T 2010 \textit{Phys. Scr.}\ \textbf{81}
025001

\noindent \lbrack 11] Oziewicz Z 2011 \textit{J. Phys.: Conf. Ser}. \textbf{%
330} 012012

\noindent \lbrack 12] Ivezi\'{c} T 2011 arXiv: 1101.3292

\noindent \lbrack 13] Ivezi\'{c} T 2012 \textit{Int. J. Mod. Phys. B}
\textbf{26} 1250040

\noindent \lbrack 14] Ivezi\'{c} T 2005 \textit{Found. Phys.} \textit{Lett.}
\textbf{18} 401

\noindent \lbrack 15] Ludvigsen M 1999 \textit{General\ Relativity}, \textit{%
A Geometric Approach}

(Cambridge: Cambridge University Press)

Sonego S and Abramowicz M A J 1998 \textit{J. Math. Phys.} \textbf{39} 3158

\noindent \lbrack 16] Minkowski H 1908 \textit{Nachr. Ges. Wiss. G\"{o}%
ttingen} 53

Minkowski H 1910 \textit{Math. Ann. }\textbf{68} 472

Saha M N and Bose S N 1920 \textit{The Principle of Relativity:}

\textit{Original Papers by A. Einstein and H. Minkowski} (Calcutta:

Calcutta University Press) (Engl. Transl.)

\noindent \lbrack 17] Vanzella D A T 2013 \textit{Phys. Rev. Lett.} \textbf{%
110} 089401

\noindent \lbrack 18] N\'{u}\~{n}ez Y\'{e}pez H N, Salas Brito A L and
Vargas C A 1988 \textit{Revista Mexicana}

\textit{de F\'{\i}sica} \textbf{34} 636

Esposito S 1998 \textit{Found.\ Phys.}\ \textbf{28} 231

Anandan J 2000 \textit{Phys. Rev. Lett.} \textbf{85} 1354

M\o ller C 1972 \textit{The Theory of Relativity} 2nd ed. (Oxford: Clarendon
Press)

Hillion P 1993 \textit{Phys. Rev. E} \textbf{48} 3060

\noindent \lbrack 19] Wald R M 1984 \textit{General Relativity} (Chicago:
The University of Chicago

Press)

Vanzella D A T, Matsas G E A and Crater H W 1996 \textit{Am. J. Phys.}

\textbf{64 }1075

Hehl F W and Obukhov Yu N 2003 \textit{Foundations of Classical }

\textit{Electrodynamics: Charge, flux, and metric} (Boston: Birkh\"{a}user)

\noindent \lbrack 20] Ivezi\'{c} T 2008 \textit{Fizika A} \textbf{17} 1;
2006 arXiv: physics/0607189

\noindent \lbrack 21] Ivezi\'{c} T 2003 \textit{Found. Phys.} \textbf{33}
1339

\noindent \lbrack 22] Ivezi\'{c} T 2005 \textit{Found. Phys.} Lett. \textbf{%
18 }301

\noindent \lbrack 23] Ivezi\'{c} T 2005 \textit{Found. Phys.} \textbf{35}
1585

\noindent \lbrack 24] Ivezi\'{c} T 2007 \textit{Phys. Rev. Lett.} \textbf{98}
108901

\noindent \lbrack 25] Ivezi\'{c} T 2008 arXiv: 0809.5277

\noindent \lbrack 26] Ivezi\'{c} T 2010 \textit{Phys. Scr.} \textbf{82}
055007

\noindent \lbrack 27] Hestenes D 2003 \textit{Am. J Phys.} \textbf{71} 691;

Hestenes D 1966 \textit{Space-Time Algebra }(New York: Gordon \& Breach);

Hestenes D 1999 \textit{New Foundations for Classical Mechanics }2nd edn

(Dordrecht: Kluwer)

\noindent \lbrack 28] Doran C and Lasenby A 2003 \textit{Geometric algebra
for physicists }

(Cambridge: Cambridge University)

\noindent \lbrack 29] Ivezi\'{c} T 2006 \textit{Found. Phys.} \textbf{36}
1511

Ivezi\'{c} T 2007 \textit{Fizika A} \textbf{16} 207

\noindent \lbrack 30] Jackson J D 2004 \textit{Am. J. Phys.} \textbf{72} 1484

\noindent \lbrack 31] Jefimenko O D 1999 \textit{J. Phys. A: Math. Gen.}
\textbf{32} 3755

\noindent \lbrack 32] Mansuripur M 2012 \textit{Phys. Rev. Lett.} \textbf{98}
193901;

Mansuripur M 2012 \textit{Proc. SPIE} \textbf{8455 }845512;

Mansuripur M 2013 \textit{Phys. Rev. Lett.} \textbf{110} 089405

\noindent \lbrack 33] Ivezi\'{c} T 2007 \textit{Found. Phys.} \textbf{37} 747

\noindent \lbrack 34] Ivezi\'{c} T 2012 arXiv: 1212.4684

\noindent \lbrack 35] Vanzella D A T 2013 \textit{Phys. Rev. Lett.} \textbf{%
110} 089401

Barnett S M 2013 Phys. Rev. Lett. \textbf{110} 089402;

Saldanha P L 2013 \textit{Phys. Rev. Lett.} \textbf{110} 089403;

Khorrami M 2013 \textit{Phys. Rev. Lett.} \textbf{110} 089404;

Griffiths D J and Hnizdo V 2013 \textit{Am. J. Phys.} \textbf{81} 570

\noindent \lbrack 36] Landau L D and Lifshitz E M 2000 The Classical Theory
of Fields

(Oxford: Butterworth Heinemann)

\noindent \lbrack 37] Ivezi\'{c} T 2007 arXiv: physics/0703139

\noindent \lbrack 38] Ivezi\'{c} T 2007 arXiv: hep-th/0705.0744

\noindent \lbrack 39] Weinberg S 1999 \textit{The Quantum Theory of Fields,
Volume I, Foundations}

(Cambridge: Cambridge University)

\noindent \lbrack 40] Arunasalam V 1994 \textit{Found. Phys. Lett.} \textbf{7%
} 515

\noindent \lbrack 41] Anandan J 2000 \textit{Phys. Rev. Lett.} \textbf{85}
1354

\noindent \lbrack 42] Wilson M and Wilson H A 1913 \textit{Proc. Roy. Soc.
(London)} \textit{A}\textbf{89} 99

\noindent \lbrack 43] R\"{o}ntgen W C 1888\ \textit{Ann. Phys. (Leipzig)}
\textbf{35} 264

\noindent \lbrack 44] G. Rousseaux, \textit{Europhys. Lett.} \textbf{84},
20002 (2008).

\end{document}